\documentclass[aps,pra, onecolumn,14 pts,a4paper,showpacs]{revtex4}

\usepackage{amssymb,amsmath,amsfonts,graphicx}
\usepackage{bm}
 \usepackage{bm,color}

\begin{document}

\title{
Turbulence: does energy cascade exist?}

\author{ Christophe Josserand$^1$, Martine Le Berre$^2$, Thierry Lehner$^3$ and Yves Pomeau$^4$} 
\affiliation{$^1$ Sorbonne Universit\'es, CNRS \& UPMC Univ Paris 06, UMR 7190, Institut Jean Le Rond d'Alembert, F-75005, Paris, France.
\\$^2$ Institut des Sciences Mol\'eculaires d'Orsay ISMO-CNRS Universit\'e Paris-Sud Bat. 210 91405 Orsay Cedex, France.
\\$^3$ Observatoire de Paris, LUTH, France.
\\$^4$  Department of Mathematics University of ArizonaTucson AZ 85721, USA
} 
\date{\today}
\begin{abstract}
\textbf{Abstract}

To answer the question whether  a cascade  of energy exists or not in turbulence, we propose a set of  correlation functions able to  test if there is an irreversible  transfert of energy, step by step, from large to small structures. These tests are applied to real  Eulerian data of a turbulent velocity flow, taken in the wind grid tunnel of Modane, and also to a prototype model equation for wave turbulence. First we demonstrate the irreversible character of  the  flow by using multi-time correlation function at a given point of space. Moreover the unexpected behavior of the test function leads us to connect irreversibility and finite time singularities (intermittency). Secondly we show that turbulent cascade exists, and is a dynamical process, by using a test function depending on time and frequency. The cascade shows up only
in the inertial domain  where the kinetic energy is transferred more rapidly (on average) from the wavenumber $k_{1}$ to $k_{2}$ than from $k_{1}$ to $k'_{2}$ larger than $k_{2}$. 

\end{abstract}

\maketitle

\textit{
Personal note by C.J. : It is a great honor for my colleagues and myself to contribute to this Special Memorial Issue Dedicated to Leo Kadanoff. The work presented here trace back in fact to my stay in Chicago as a post-doc with Leo, where I was investigating how irreversible dynamics could emerge from reversible systems. I remember emotionally my discussions with Leo during my postdoc where his advices were always of great help, often concluding with his characteristic voice by ''I would (not) encourage you to go in that direction''! My two years in Chicago have widely influenced my scientific activity thanks to Leo's personal advices and to the outstanding atmosphere there whose Leo was at the heart.}

\section {Introduction}

 The theory of turbulence by Kolmogorov  is based upon the idea that at very large Reynolds number the energy flows (in 3D incompressible turbulence stationary and isotropic on average) from large scale, $l_{0}$ where it is injected, to small scales, $\ell_{\nu}$ where it is dissipated by viscosity. In intermediate scales  (the so-called inertial range, $\ell_{\nu} << \ell << \ell_{0}$) the energy is assumed to be transferred  without any dissipation step by step from large  to smaller scales by an inviscid cascade. We believe that the cascade hypothesis
 is yet to be directly verified, that still remains a current challenge. From the cascade assumption, Kolmogorov derived scaling laws which have been widely tested  both numerically and experimentally. Because of their formal simplicity, those scaling laws cannot be seen, we believe, as checking every aspect of the theory of cascade. They rely on a general assumption of dependence with respect to a physical parameter, the power dissipated per unit volume,  formally without any other assumption. 
 
 The scaling laws derived by Kolmogorov in his first paper K41 \cite{K41} are deduced from the hypothesis of constant energy flux (or constant rate of  energy transfer, $\epsilon$, the rate of injection of the energy) at each length scale $\ell$. This gives  a velocity at scale $\ell$ of order of magnitude 

 \begin{equation}
 u_{\ell} \sim  ( \ell \epsilon )^{1/3} 
 \textrm{,}
\label{eq:ul}
\end{equation}
 and  a time duration  of fluctuations at this scale of order 
  \begin{equation}
   \tau_{\ell} \sim  (\ell^{2}/\epsilon )^{1/3}
   \textrm{.}
\label{eq:taul}
\end{equation}
 Later, in K62 \cite{K62}, Kolmogorov derived a statistical model, also based on the cascade hypothesis including fluctuations (observed in time records as an alternation of quiet and bursts phases)  where it turns out that simple scaling laws are absent. In these description the transfer of energy from one scale to a smaller one occurs  randomly via instabilities  of the smaller scales taking some time, of order $\tau_{\ell}$.
 Among the many works following Kolmogorov masterpiece, let us mention the recent paper by Ruelle \cite{Ruelle} where hydrodynamical turbulence is reformulated as a heat flow problem
through a collection of coupled Hamiltonian systems with weak local interactions,
 of smaller and smaller extent (keeping therefore the cascade idea). Assuming local Boltzmann equilibrium,  the author obtains corrections to the original scalings of K41, and goes beyond  the results of K62 (log-normal distribution of the radial velocity) by including intermittency, while maintaining the property of  \textit{time delay depending on the size of the structure}. 
 
 The present work is an attempt to show by analyzing  experimental data that the energy present at large scales is effectively transferred to smaller scales in a dynamical process, requiring therefore a finite time to go from a given scale to a smaller one. This process takes some time and is fundamentally irreversible. This is why we put it in evidence by looking  
 at pertinent time correlation functions. This proves first the irreversibility of the turbulent fluctuations as occurring in an out-of-equilibrium stationary system,  and secondly we measure the time delays of the transfer of energy between the different scales $\ell$. Starting from a given "large" scale, this delay increases as one goes to smaller and smaller scales: there is a priori no exchange of energy between a large scale and a much smaller scale, which is only convected by the velocity fluctuation at large scale. Therefore the "elementary" transfer of energy must be from a given scale to a smaller one, but not much smaller and the feeding of very small scales can be done only step by step in the cascade process. As written above,  the dependence of the time delay with respect to the size of the structure is a signature of the Kolmogorov cascade as a dynamical process. Were the description of  turbulence in terms of cascade incorrect, the time of transfer of energy should not depend on the size of structures in a wide range of length scales. For example no dynamical cascade of energy  were necessary to transfer energy to small scales if the  solution of the Navier-Stokes or Euler equations displays a finite time singularity, by analogy with what happens for instance with Burgers equation, something already suggested by Leray \cite{Leray} to explain turbulence. In this case the transfer of energy from large initial structures to small eddies should occur in a single step on a time scale depending only on the initial data (something testable in principle), therefore  the  rate of energy transfer should be the same for large and intermediate scales.  
 
 We present below applications of those ideas to real physical data of turbulent flow and also to a model of wave turbulence well known to display  irreversibility and cascade. Experimental Eulerian data are generally used in the frame of Taylor's frozen turbulence hypothesis which assumes that turbulent fluctuations are carried with the mean flow in a quasi-frozen manner \cite{Taylor} . This hypothesis avoids multi-point measurements  which turn out to be a severe challenge. The spatial velocity field is then reconstructed from single (or few) discrete pointwise probes. This yields time dependent signals that are then mapped into spatial domain. The 
frequency $\omega$ is related to the streamwise wavenumber $k$ by the simple relation 
\begin{equation}
\omega=kv_{0}
\textrm{,}
\label{eq:taylor}
\end{equation}

 where $v_{0}$ is the mean velocity of the flow. The validity of Taylor's hypothesis  is discussed in \cite{Taylor-val}. There are examples of turbulent flows, like the von Karman flow between counter rotating discs  without mean velocity field at well defined locations where one can measure in principle actual time dependent correlations, however they are far from homogeneous and so not considered here. 

We point out that the data we deal with concern Eulerian velocity fields, not Lagrangian ones.  Such Eulerian velocity fields are the relevant ones to test Kolmogorov ideas on turbulent cascades and time irreversibility in turbulent flows, even though some numerical simulations of the fluid equations are done in the Lagrangian framework, see \cite{Toschi}. In a recent  paper Jucha et al.\cite{Jucha} also studied  the problem of irreversibility in turbulence by using Lagrangian data. In this framework, the ongoing stretching of the convected material points yields a fake time dependence: it would exist even for a steady flow. What is measured by following Lagrangian trajectories is an inextricable mixture of the randomness of the flow structure and of its own dynamics. 
In our opinion the idea of  transfer of energy across the scales cannot be tested this way, because the dynamics of eddies is measured in \cite{Jucha} by the evolution of the spatial separation between two particles.  

On the contrary we shall derive  the time delay characterizing the transfer of energy from one scale to a smaller scale from our analysis of the \textit{Eulerian data}. To put in evidence this delay, we have to look  in principle at the whole time and space dependence of some test functions associated to irreversibility, as already suggested in ref.\cite{pom82}. This should require 
to have experimental data giving access to  time dependent fluctuations at several different locations, or to make Taylor's hypothesis. But the multi-spatial datas are hardly accessible in regular wind tunnel turbulence, the most studied flow for checking Kolmogorov theory. We had at our disposal data obtained in the nineties from the wind tunnel of Modane (the largest wind tunnel in Europe located in the French Alps), which consist
in  full temporal series taken from  a single detector (hot wire) placed at a given point in the turbulent air flow. Therefore we had to  make Taylor hypothesis. It  amounts to assume that the flow is frozen, the correspondence between time and space being given by relation (\ref{eq:taylor}).  Within this frame we find a positive irreversibility test and a positive cascade test.  In addition we find an unexpected behavior of the irreversibility test function at small time, which is due to a
 lack of smoothness of the turbulent signal, something related in general to what is called intermittency. In the case of Modane's data we interpret the test function behavior by assuming that it is due to the recording of pointwise singularities passing randomly near the probe.  Comparing the short time behavior of the test function with the long range statistics of the acceleration, we conclude that the impact of finite time singularities on the velocity flow  is more complex  that the one usually assumed, a point discussed in subsection  \ref{sec:sing-accel}.
 
Furthermore  we also investigate those properties of irreversibility and cascade  using a prototype model  for wave turbulence (with one spatial dimension) first introduced in ref.\cite{MMT}. This model is a partial differential equation in one space dimension for a function depending on time. In the linear approximation it yields waves with the same dispersion relation as water waves.  The nonlinear term introduces wave interactions which can be described in the limit of small but not vanishingly small amplitudes by the equation of wave turbulence. Those equations describe an energy transfer from large to small scales with a spectrum of constant energy flux analytically predicted under conditions of small nonlinearity.  Direct numerical simulations allow us to have access to local  and global  spatio-temporal quantities, that are used to show how this irreversible transfer of energy shows up in time correlation functions testing the breaking of symmetry of time reversal invariance. A preliminary study of the cascading transfer of energy is also presented.  

We first present  our study of  irreversibility (section \ref{sec:irreversibility})  and turbulent cascade (section \ref{sec:cascade}) to  experimental Modane data, then we apply our test functions to the spatio-temporal model  of wave turbulence (\ref{sec:wave-turb}). This study deals exclusively with signals assumed to be stationary at least in wide sense (with mean independent of time and autocorrelation dependent on the difference of times).

\section {Test  for irreversibility in general and for turbulent flows}
\label{sec:irreversibility}

Consider a  stationary random  signal $x(t)$ (it depends on time but has statistical properties independent of time), which can be the records of a hot-wire probe in a turbulent flow or the scintillation of light emitted by a excited atom or many other examples. As Onsager had shown  \cite{Onsager}, by measuring such a signal there is no way to make a difference between the two possible directions of time if the signal comes from a system at equilibrium like a black-body radiation for instance. But there are many examples of systems which are out-of-equilibrium. Therefore measurements of their time fluctuations should permit to make a distinction between the two possible directions of time and give access to the origin of  dissipation in such systems. This kind of irreversibility cannot be put in evidence by looking at auto-correlation function of the form $\Gamma_{x}(t,t+\tau)= <x(t)x(t+\tau)>$ , as soon as $x(t)$ is the fluctuation of a real ergodic process, because in this case  this quantity  is insensitive to the  direction of time, since $\Gamma_{x}(t,t+\tau)= \Gamma_{x}(\tau)= \Gamma_{x}(-\tau)$. 

Other auto-correlation functions  \cite{pom82}  may display a difference between the two directions of time. Even more it is possible to define auto-correlation functions which are exactly zero if the system is time reversible and which do not vanish if it is not invariant under time reversal symmetry. 

Let us define test functions allowing to test the symmetry of the fluctuations under time reversal. To test the irreversible character of the turbulent signal  we shall use  the following third order correlation functions,
\begin{equation}
 \Psi_{1}(\tau) = < x(t)\; [x(t+2\tau) - x(t+\tau)] \; x(t + 3\tau)>
\textrm{,}
\label{eq:function.3t}
\end{equation}
and
\begin{equation}
 \Psi_{2}(\tau) = < x^{2}(t)\; x(t+\tau) > -  < x(t)\; x^{2}(t+\tau) >
\textrm{.}
\label{eq:psi2}
\end{equation}
Both correlation functions change sign by reversing time.  They are odd functions of $\tau$,  and the Taylor expansions  near $\tau = 0$ begin with a cubic term both for for $\Psi_{1}$ and $\Psi_{2}$,  at least if some conditions are satisfied by the fluctuations of the derivative of  $x(t)$ (namely the acceleration), which seemingly do not happen for the turbulent velocity. To show this point, let us compute the first two non zero terms of the Taylor expansion of   $\Psi_{2}(\tau)$ near $\tau = 0$. After shifting $t$ of $(-\tau)$ in $-  < x(t)\; x^{2}(t+\tau)>$ one obtains 
\begin{equation}
 \Psi_{2}(\tau) = < x^{2}(t)\ ( x(t+\tau) - x (t-\tau)) > = 2  \tau <  x^2 \dot{x}(t) > + \frac{\tau^3}{3}  <x^2 \dddot{x}(t)>  + ...
\textrm{.}
\label{eq:psi2T}
\end{equation}
Because $x^2 \dot{x}(t) $ is the  time derivative of $x^3/3$ its mean value should be zero. The same argument does not apply to the  coefficient of $\tau^3$, namely $<x^2 \dddot{x}(t)>$ which is not the average value of a time derivative. However, as we shall see below the function $ \Psi_{2}(\tau) $, when computed from the fluctuations of turbulent velocity in the wind tunnel of Modane, does not have at all the property of decaying to zero at $\tau = 0$ with a cubic law.  On the contrary the behavior of  $\psi_{2}(\tau)$ at small time is  well fitted by a linear time dependance, see below (Fig.\ref{fig:tau small}).  Examples of functions everywhere continuous but differentiable almost nowhere exist, like the 
 Weierstrass function. 
 For the turbulent velocity field we are tempted to associate the non differentiability of the function $x(t)$, $x$ being a velocity, $ \dot{x}(t)$ is an acceleration,
to the observation already made  \cite{Acceleration} that the probability distribution of the acceleration in fully turbulent flows decreases slowly, displaying noticeable contribution for large acceleration values, and could even be non normalizable. This could result from the random occurrence of point singularities of the velocity in space and time, as considered in  \cite{CRASYP}.  Because that could explain the behavior of the test function $\Psi_{2}$ close to zero observed in Modane's data, we  detail this point in the next subsection.
 
\subsection{Singularities with Sedov-Taylor  and incompressible Euler scalings}
\begin{figure}
\centerline{
 \includegraphics[height=1.5in]{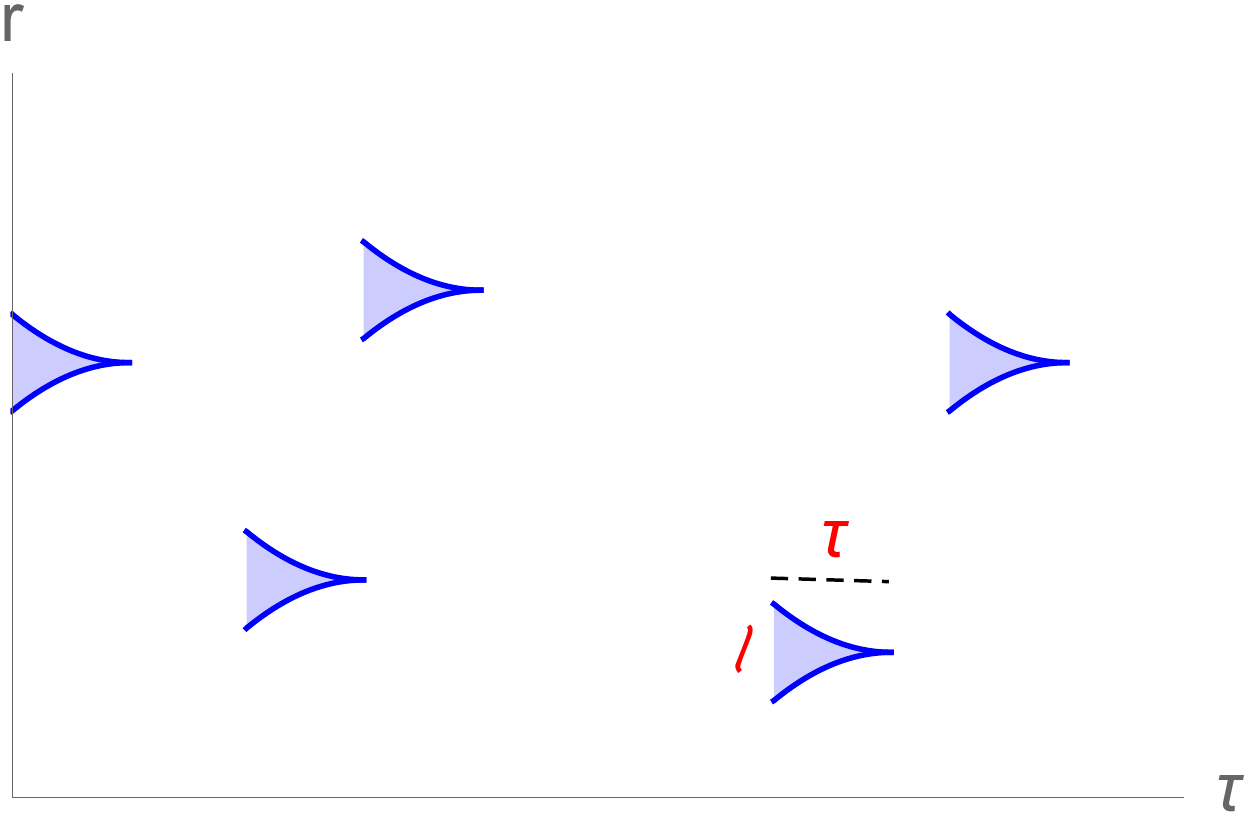}
  }
\caption{
Schematic representation of space-time volumes (in one dimensional space) close to singular points assumed to be uniformly distributed in $R\times R$. 
}
\label{fig:sing}
\end{figure}

In this subsection the velocity is labelled as $v(t)$.   Let us assume that  $v(t)$ becomes singular at some times $t_{i}$ and some spatial points $y_{i}$, with a power law of the form 
 \begin{equation}
v(t) \sim \tau^{-\alpha}
\textrm{,}
\label{eq:expv}
\end{equation}
where $\tau= t_{i}-t$ is   the positive time lag between $t$ and $t_{i}$.
That relation implies that the spatial dimension of a singularity is of order 
 \begin{equation}
\ell \sim \tau^{1-\alpha}
\textrm{,}
\label{eq:expy}
\end{equation}
and the space-time volume of the singularity is 
\begin{equation}
Q(\tau)=\tau \ell^{3}(\tau)
\textrm{,}
\label{eq:vareps}
\end{equation}
for a time interval of order $\tau$ from the singular point, see Fig.\ref{fig:sing}.  Assuming a uniform distribution of  the singular points $t_{i}, y_{i}$, the function $Q(\tau)$ becomes the  probability distribution of being at a distance $\tau$ from a singular point in $R^{4}$ (up to a norm factor independent of $\tau$), in other words $Q(\tau)$ is the  probability  for the detecting device to be inside a fluctuation yielding a point-wise singularity occurring at time  $\tau$ afterwards. In Sedov-Taylor theory\cite{SedovTaylor} originally derived for blast waves, the velocity would scale as
 \begin{equation}
v  \approx  E^{1/5} (t^* - t)^{-3/5}
\textrm{,}
\label{eq:sedov}
\end{equation}
 where $E$ is the energy focused in the vicinity of the singularity and $t^*$ the time of the singularity.  Notice that the Sedov-Taylor theory has been derived originally for a blast wave initiated by an explosion of a given energy, and it is valid for long times after the initial explosion. Another example of  random occurence of singularities concerns the solutions of the Euler incompressible equations, where the same exponent is found (when assuming a finite energy), but there the time is running from negative values, before the singularity, to zero, the time of the point-wise singularity. The exponents are the same in the two cases, but the general picture is completely different.  In particular the sense of time in the Euler case and in turbulence studied here is opposite to the original Sedov-Taylor case.

\subsubsection{Singularity with finite energy}
Let us apply to our problem the method used for singular solution of the Euler equations,  which is an extended form of the  Sedov-Taylor self-similar solution. We assume  a velocity   scaling as (\ref{eq:sedov}),  and a power law behavior of the test function of the form,
\begin{equation}
\Psi_{2}(\tau) \sim \tau ^{\xi}  
\textrm{,}
\label{eq:psi2-0}
\end{equation}
close to $\tau = 0$ with $\xi$ positive. We make the hypothesis that the non cubic behavior of $\Psi_{2}(\tau)$ results from the fact that   the quantity $v(y, t+\tau) -v(y,t-\tau)$ diverge randomly in space and time  at  $(y_{i},t_{i})$.  As a consequence the acceleration $\dot{v}(t_{i})$ recorded  by the hot wire undergoes a peak when the singular point is approaching, this peak disappearing afterwards.  

We may estimate 
the average $ \Psi_{2}(\tau)=<v^2 (t)[v(t+\tau) - v(t-\tau)]>$ 
according to scales defined by the Sedov-Taylor singularity (and later compare with our observations).  The length scale (\ref{eq:expy}) is of order 
\begin{equation}
\ell_{s}(\tau) = \tau^{2/5} 
\textrm{,}
\label{eq:ells}
\end{equation}
 for a time interval of order $\tau$. To estimate $\Psi_{2}(\tau)$ at small $\tau$, we first consider
the quantity $ v^2 (t)[v(t+\tau) - v(t-\tau)]$  to be averaged in each space-time volume. That quantity  is  of order $ v^{3}  \sim (\ell_{s}/\tau)^{3}$   because in the vicinity of the singularity we can replace the velocity $v$ by $\ell_{s}/\tau$.  From (\ref{eq:vareps}), the probability to be at distance $\tau$ from a singularity is $Q_{s}(\tau)=\tau \ell_{s}^{3}(\tau)$, which is the contribution of the Sedov-Taylor singularities to the latter quantity. 
Therefore Sedov-Taylor  singularities  would give the following averaged value  
\begin{equation}
 \Psi_{2}(\tau)=Q(\tau) (\ell_{s}/\tau)^{3}
\textrm{.}
\label{eq:vareps-sedov}
\end{equation}
for the test function near $\tau = 0$, which is of order  $ \ell_{s}^{6}/\tau^{2}$ or $\Psi_{2,s}(\tau) \sim  \tau^{2/5}$. That gives 
\begin{equation}
\xi_{s}= \frac{2}{5}    \qquad  \text{for}  \qquad  \alpha_{s}=\frac{3}{5}
\textrm{.}
\label{eq:exp-sed}
\end{equation}
 As we shall see this exponent is too small to represent the data, which display 
$\Psi_{2}(\tau)$ linear with respect to $\tau$ near $\tau = 0$, and correspond to a case considered in the next subsection, see equation (\ref{eq:exp-Modane}). 

\subsubsection{Singularities with decaying energy} 
\label{sec:sing-psi}
In Sedov-Taylor theory of blast-waves, the energy $E$ in the collapsing volume remains finite, see equation (\ref{eq:sedov}). 
However  a singular solution of incompressible Euler equation in 3D does not necessarily has a finite (non zero) energy in the collapsing volume. It could happen that the velocity becomes singular at a point, but with no energy stored in the singular domain, more precisely with energy  decaying to zero  with a positive power of $\tau$.  Such a situation occurs for singularities of the nonlinear Schr\"odinger equation in three dimensions of space and in the focusing case, where the energy in the collapsed region tends to zero at collapse. 
Now we assume  that the velocity becomes singular with a power law of time, equation (\ref{eq:expv}),  where $\alpha$ is an (yet unknown) exponent.  The size  of the collapsing region is  still given by (\ref{eq:expy}). The energy $E \sim  v^{5}\tau^{3}  \sim \tau^{(3-5\alpha)}$, see relation (\ref{eq:sedov}),  is  finite for $\alpha =3/5$  (the Sedov-Taylor case treated above), or  tends to zero with $\tau$  if $0 <\alpha< 3/5$. Using the same arguments as before, namely the relation $ \Psi_{2}(\tau)\sim  \ell^{6}/\tau^{2}$ one finds  $ \Psi_{2}(\tau) \sim \tau^{4 -6\alpha}$ or,
\begin{equation}
\xi= 4 -6\alpha
\textrm{.}
\label{eq:exp-other}
\end{equation}
The condition for $\Psi_{2}(\tau)$ to have an infinite derivative at $\tau = 0$ but tends to zero with $\tau$ is fulfilled for any 
\begin{equation}
1/2 <\alpha < 2/3
\textrm{,}
\label{eq:euler}
\end{equation}
 an interval including the Sedov-Taylor value $3/5$.  This leaves a rather narrow window for the exponent $\alpha$, $5/10 <\alpha <6/10$.  The 
lower bound  of (\ref{eq:euler}) corresponds to  linear behavior of $\Psi_{2}(\tau)$  (finite derivative) at $\tau=0$,  which is observed numerically in the study of Modane's data, see below, 
\begin{equation}
\xi=1     \qquad \text{for}    \qquad  \alpha= \frac{1}{2} 
\textrm{.}
\label{eq:exp-Modane}
\end{equation}

We present now an explanation for this value of $\alpha$, based upon a highly non trivial property of the Euler equation for incompressible flows and on the assumption of existence of finite time singularity described by a self-similar singular solution of the Euler equation in 3D. We already pointed out the possibility that energy conservation inside the collapsing region may not be necessarily valid (there is in this case a leakage of energy outside the collapsing domain) which allows in principle other exponents than the Sedov-Taylor ones. However there are other constraints on the exponent of self-similarity arising from the properties of the Euler equation in 3D. One of those properties is the Kelvin theorem of conservation of the circulation of the fluid velocity along a closed curve carried by the flow, $\int{v ds}$ with $s$ the  coordinate along the curve. In a self similar solution such a closed curve is transported by a velocity field of amplitude growing like (\ref{eq:expv}), with a length scale, as any other length scale inside the collapsing domain,  decreasing like (\ref{eq:expy}), $\alpha$ being less than $1$. Therefore the circulation is conserved in the course of time until the
time of the collapse if $\alpha = 1/2$, which is the value explaining the
slope of $\Psi_{2}(\tau)$ near $\tau=0$.  From the conservation of the
circulation in the collapse domain, we deduce that  the local Reynolds
number (in time and space), $Re_{loc}$, which is of order of the
circulation divided by $\nu$, is also constant in this domain.
Therefore this local  number is
a free parameter for the similarity solution depending on the initial
conditions. Obviously 
a self-similar solution could exist at large value of $Re_{loc}$ only
(or perhaps until a time $\tau$ depending on this initial Reynolds number).
We also remark that for $\alpha = 1/2$ the
equation for the self-similar solution makes the viscosity term in
Navier-Stokes equation of the same order of magnitude as the other terms
in the limit $\tau$ tending to zero. It means somehow that the Reynolds
number in the collapsing domain remains constant.

\subsubsection{Behavior of $\Psi_{2}(\tau)$  at small  $(\tau)$ and large fluctuations of the acceleration} 
\label{sec:sing-accel}

We would like to make more precise the connection between the behavior of $\Psi_{2}(\tau)$ near $\tau=0$ and the observed large fluctuations of the acceleration in turbulent flows. The experiments measured directly the acceleration of a Lagrangian particule although we are concerned here with the time derivative of the velocity measured at a given location. It is not totally obvious that the occurrence of a large Lagrangian fluctuation is related to a large time derivative of the  Eulerian velocity. As we have no access to the Lagrangian acceleration in the conditions of the wind tunnel of Modane, it is not necessary to speculate whether the two observations (large Lagrangian acceleration and large time derivative of the Eulerian velocity) are connected or not. Here we consider the problem within the Eulerian data of the wind tunnel. 

More specifically we outline a derivation of the statistics of large deviations of the acceleration $\gamma$ (understood again as the time derivative of the Eulerian velocity) by using the same idea as above of finite time singularities of solutions of the Euler equation  with a single definite exponent $\alpha$.  Starting from the probability $Q(\tau)$, equation (\ref{eq:vareps}),  for the detecting device to be separated by the time lag $\tau$ from a singular point, we are able to derive 
 the behavior of the probability $P(\gamma)$  for large $\gamma$. Using (\ref{eq:expv}), the order of magnitude of the acceleration near the singularity is  $\gamma (\tau) \sim \tau^{- \alpha - 1}$, and the power law for the probability  $P(\gamma)$ is derived from $P(\gamma) = Q(\tau)  \frac{d  \tau}{ d  \gamma}$, one obtains after a little algebra the behavior of $P(\gamma)$ at large $\gamma$,
\begin{equation}
P(\gamma)  \sim \gamma^{\frac{2\alpha - 4}{\alpha+1}}\textrm{.}
\label{eq:gamma-exp}
\end{equation}
For $\alpha=1/2$, the distribution (\ref{eq:gamma-exp}) gives an asymptotic behavior as $\gamma^{-2}$ which does not correspond to the law deduced from Modane's data. Let us develop this point.

A self-similar solution  is usually supposed to be of the form  $v^{(0)}(r,t)=t^{-\alpha}V(r/ t^{(1-\alpha)})$ close to a singularity located at $(0,0)$, as we assumed above. If one adds the constraint of conservation of circulation, it follows that the unique exponent should be
 $\alpha = 1/2$. 
But, as argued in ref \cite{CRASYP}, a singular solution of the Euler equations could have a dependence with respect to time more complex than $v^{(0)}$ which involves a single exponent. 
The idea amounts  to change the original variables $(r; t)$ into new ones, which should be here  $(r/t^{1/2}; \; \lambda(t))$,  the velocity becoming of the form $t^{-1/2}V(r/t^{1/2}; \; \lambda(t))$. 
In the case $\lambda(t)=\log(t)$, the equations are autonomous with respect to $\lambda$  and the solution doesn't depend on $\lambda$.  But other kinds of solution can also exist with a more complex behavior at  $\lambda$ infinite  ($t\to 0$  at singular point), like oscillations or growth with an exponential of a power of $\lambda$ which is less than one, for instance $v(r,t) =\exp(\lambda^{1/2}) \; v^{(0)}(r,t)$. This last case is interesting because, without changing the exponent in $\Psi_2 (\tau)$ it could explain the observed changing slope of the probability distribution of the acceleration at large values.  We plan to return to this problem in future.

\subsection {Test  for irreversibility for trial random functions}
\label{sec:irrev-trial}
\begin{figure}
\centerline{
 \includegraphics[height=1.5in]{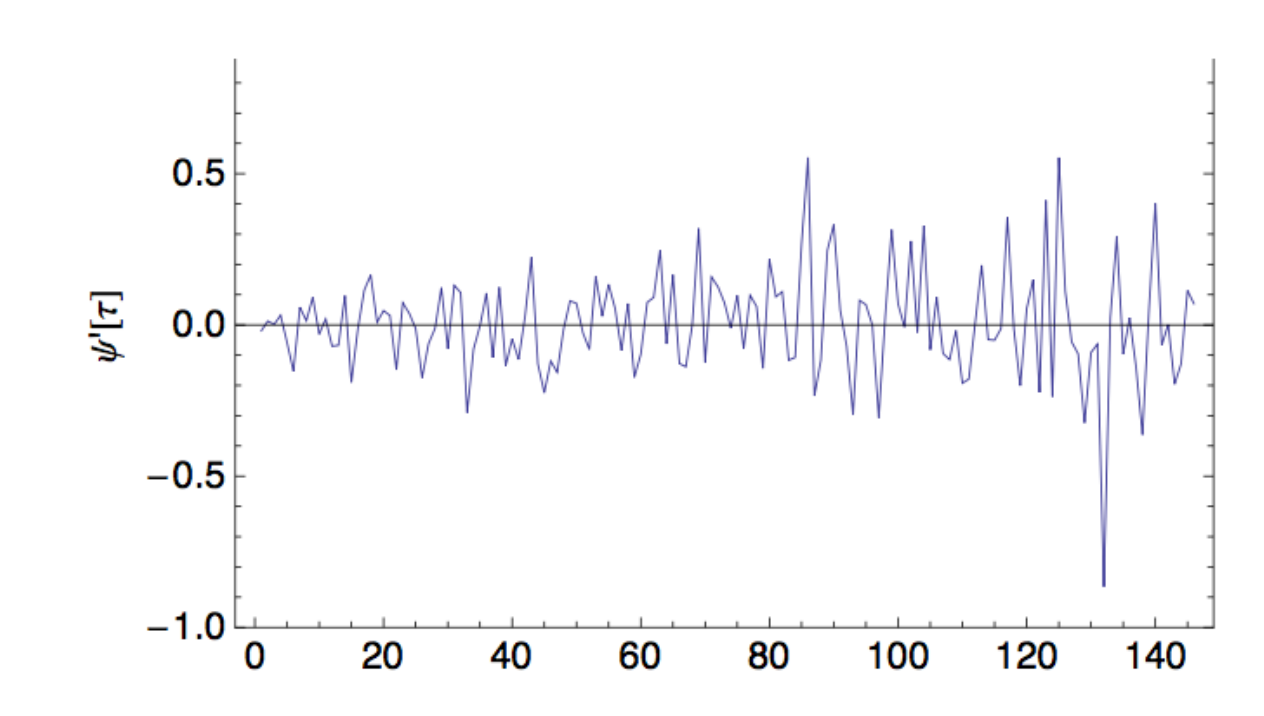}
  }
\caption{
Test $\Psi_{1}$ for a $\delta$-correlated Gaussian  noise. 
}
\label{fig:trial1}
\end{figure}

The test function $\Psi_{1}$ is shown in Fig.\ref{fig:trial1} for a typical Gaussian random walk $x(t)$ motion with initial condition seed. Clearly this noisy function remains oscillating around zero, that is the signature of a \textit{time-reversible} motion, as expected for such a case with no memory (zero correlation time). 

\subsection {Test  for irreversibility in Modane experiment}
\label{sec:irrev-Modane}

\begin{figure}
\centerline{
\includegraphics[height=1.75in]{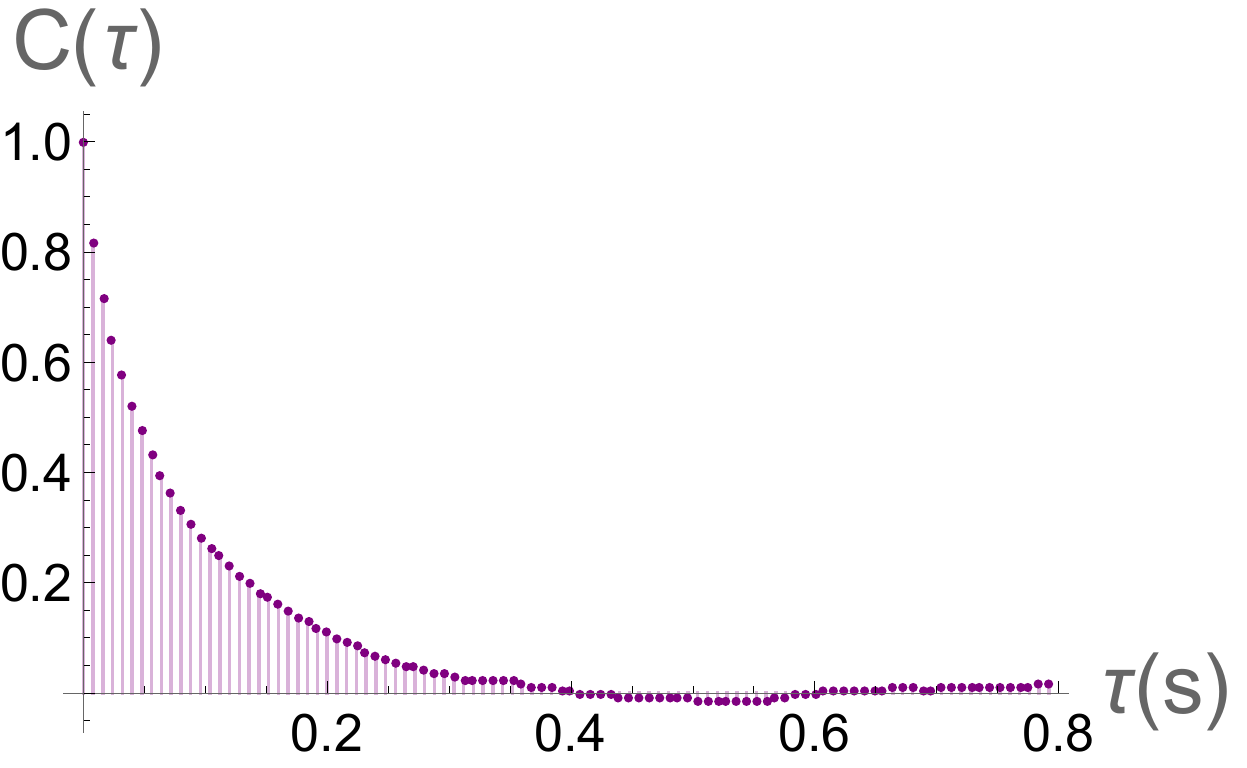}
  }
\caption{Correlation function $C(\tau)$ for the Modane experiment.
}
\label{fig:correl}
\end{figure}

This subsection is to show that the turbulent velocity fluctuations of the Modane wind tunnel do lack time reversal symmetry, which is already a non trivial result. 
We have considered time 
 series of Eulerian velocities $v(t)$  taken at one spatial point  in the experiment performed by Y. Gagne et al \cite{gagne} in the Modane  wind  grid tunnel. Details of the experimental set-up can be found in Ref.\cite{modane}. The measurements were done with  hot wires, assuming King's law for the voltage calibration.  We recall shortly the  conditions of this  experiment : the Reynolds number  $Re_{\lambda}=\sqrt{15Re}$ is equal to 2500, so that the regime is in fully developed turbulence. The measurements were made in the return vein of the wind tunnel, where turbulence is not really isotropic, but mainly resulting from the separation of an unstable boundary layer. The sampling time is $t_{s}= 1/25 ms$ ($f_{s} = 25 khz$). It is smaller than the dissipation time $ t_{\eta}=1/10.7 ms$.  As written above, we assume ergodicity of the velocity flow, it follows that any average is calculated as a running time over the full  data which consists in $210$ files of $2^{16}$ points each, that gives a total record time of about $10$ minutes. 
The average velocity is $v_{0}= 20.53 m/s$, and the standard deviation is $\sigma_{v}= \sqrt{<(v-v_{0})^{2}>}= 1.68m/s$. The correlation function of the velocity, defined by
 \begin{equation}
C(\tau)= \frac{<v(t)v(t+\tau)> - v_{0}^{2}}{\sigma_{v}^{2}}
\mathrm{,}
\label{eq:cort}
\end{equation}
 is shown in Fig. \ref{fig:correl}. It displays  a long tail with very small amplitude oscillations until $\tau \sim 0.8s$, and  a narrow central peak of half-height width equal to $\tau_{c}=45ms$.

\begin{figure}
\centerline{
 \includegraphics[height=1.5in]{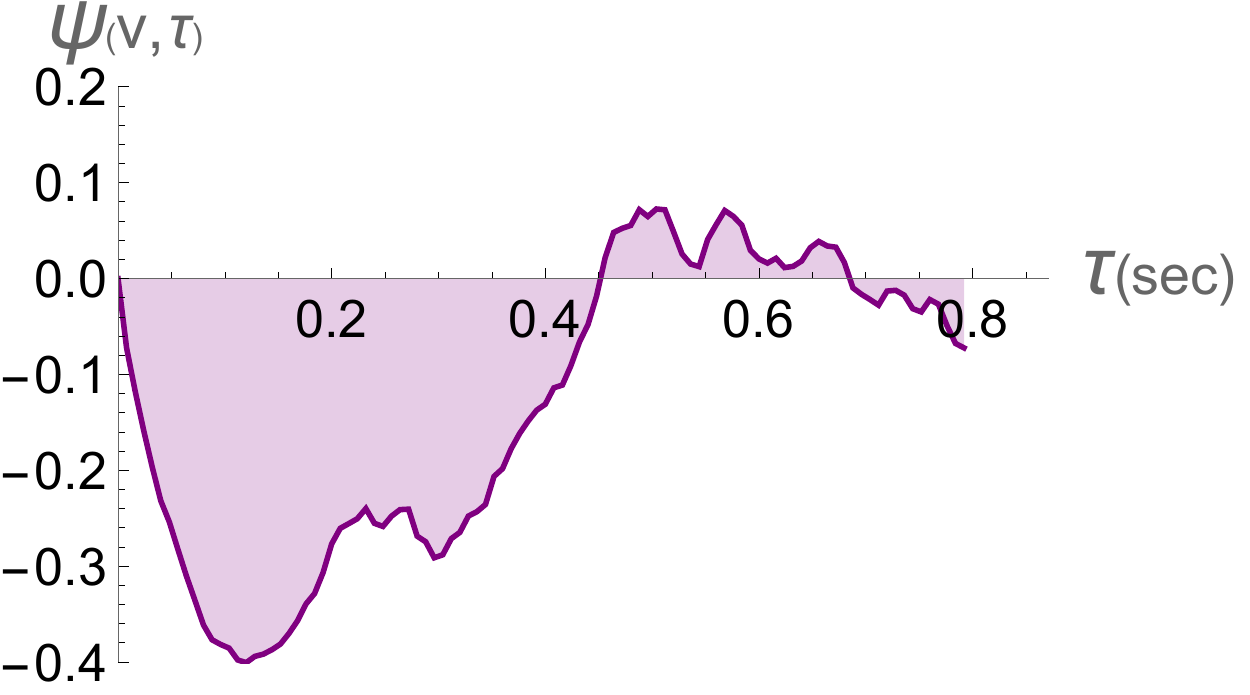}
  }
\caption{
Irreversibility test for Eulerian datas of Modane experiment,  $ \Psi_2(\tau)$ defined by equation (\ref{eq:psi2}) with $x(t)=v(t)-<v(t)>$.
}
\label{fig:test-mod-irrev}
\end{figure}

To investigate  irreversibility  in the Eulerian data of Modane experiment, we use the test function $ \Psi_{2}(\tau)$ with $x(t)=v(t)-v_{0}$. 
For a stationary signal, assuming that its average exists,  one expects that  $\Psi_{2}(\tau)$ vanishes at large time, because the variables $x(t)$ and $x(t+\tau)$  are there  uncorrelated. 
 This is what is observed in  Fig.\ref{fig:test-mod-irrev} where the  function $\Psi_{2}(\tau)$ vanishes  for $\tau\sim 0.8s$, in agreement with the behavior of $C(\tau)$. 
  
\begin{figure}
\centerline{
\includegraphics[height=1.75in]{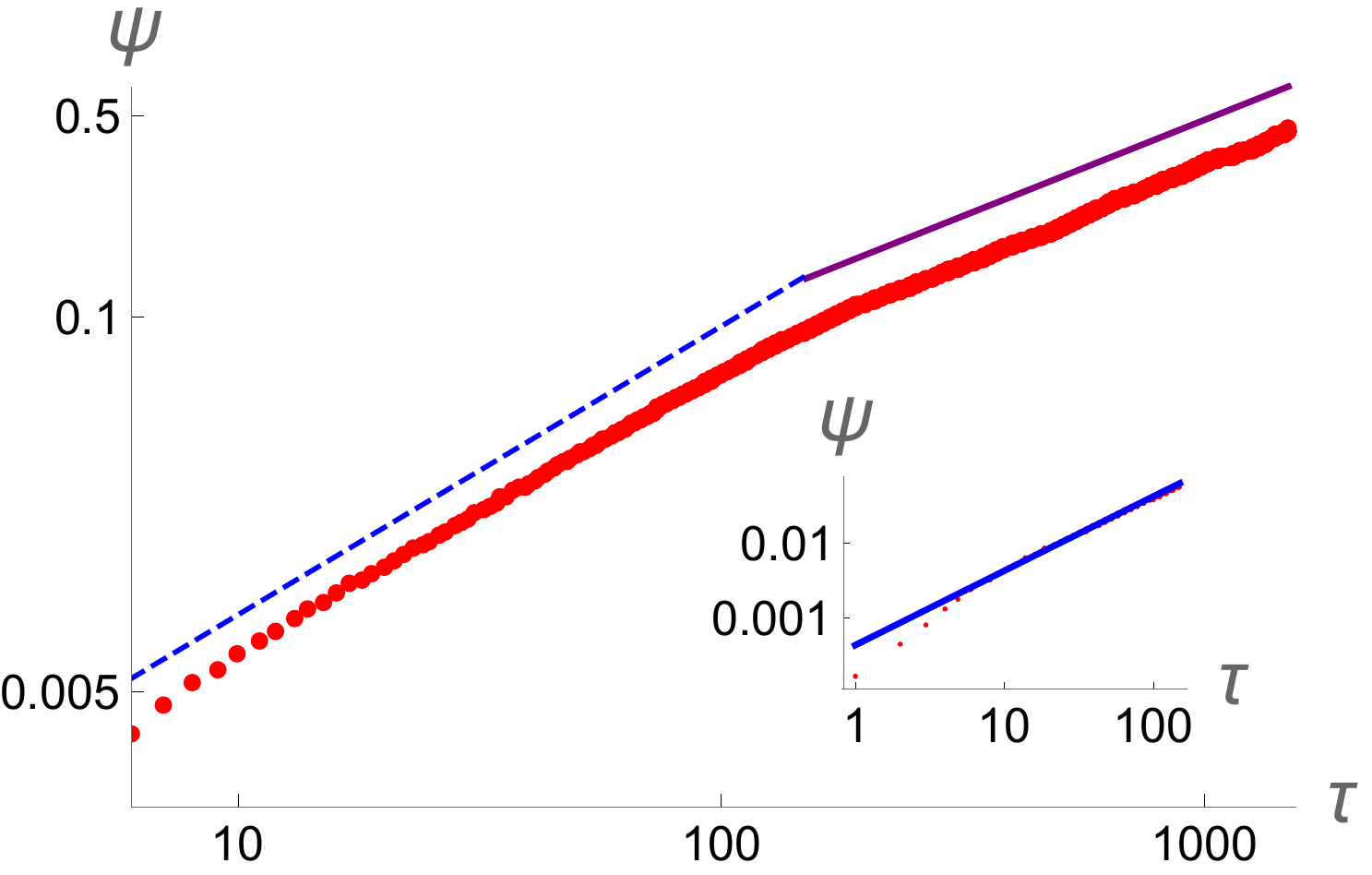}
  }
\caption{ Log-scale behavior of the function $-\Psi_{2}$ for the Modane experiment. Time is scaled to $t_{s}$,  $\Psi_{2}$  in a.u. 
The inset is  for $\tau$  close to zero, the straight blue line has slope unity.  The main curve is  for a longer time  interval , $60$ms, which corresponds to half-height of the extremum of $\Psi_{2}(\tau)$,  dashed line with slope unity,  solid line with slope  $2/3$.  }
\label{fig:tau small}
\end{figure}

For very small times $\Psi_{2}(\tau)$  grows linearly, as shown  in Fig.\ref{fig:tau small}.  Note that this behavior concerns a  time domain of order $\tau_{c} \sim 4$ms, much smaller than the width of the correlation function. For larger time values (in the growing stage of $\Psi_{2}$), the curve bent down, as indicated in the captions of 
Fig.\ref{fig:tau small}.

\begin{figure}
\centerline{
(a)\includegraphics[height=1.5in]{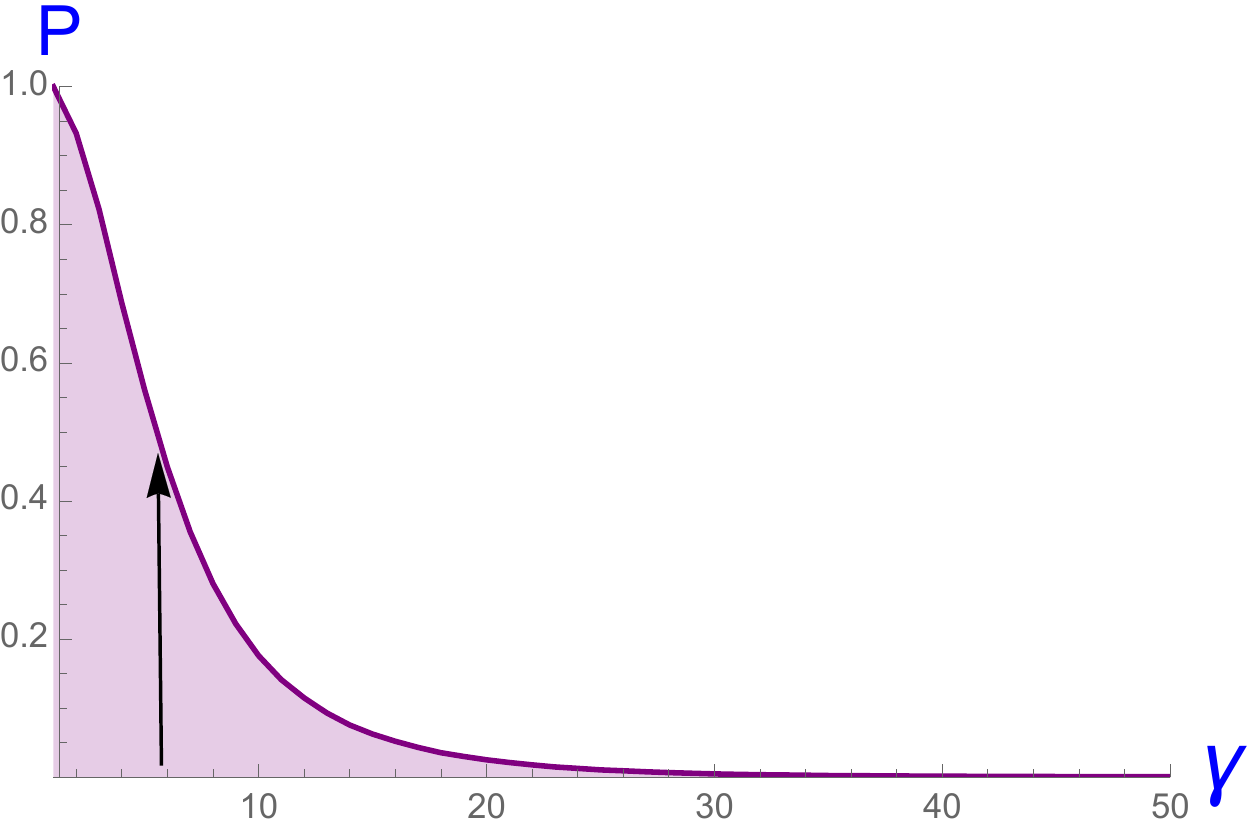}
(b)\includegraphics[height=1.75in]{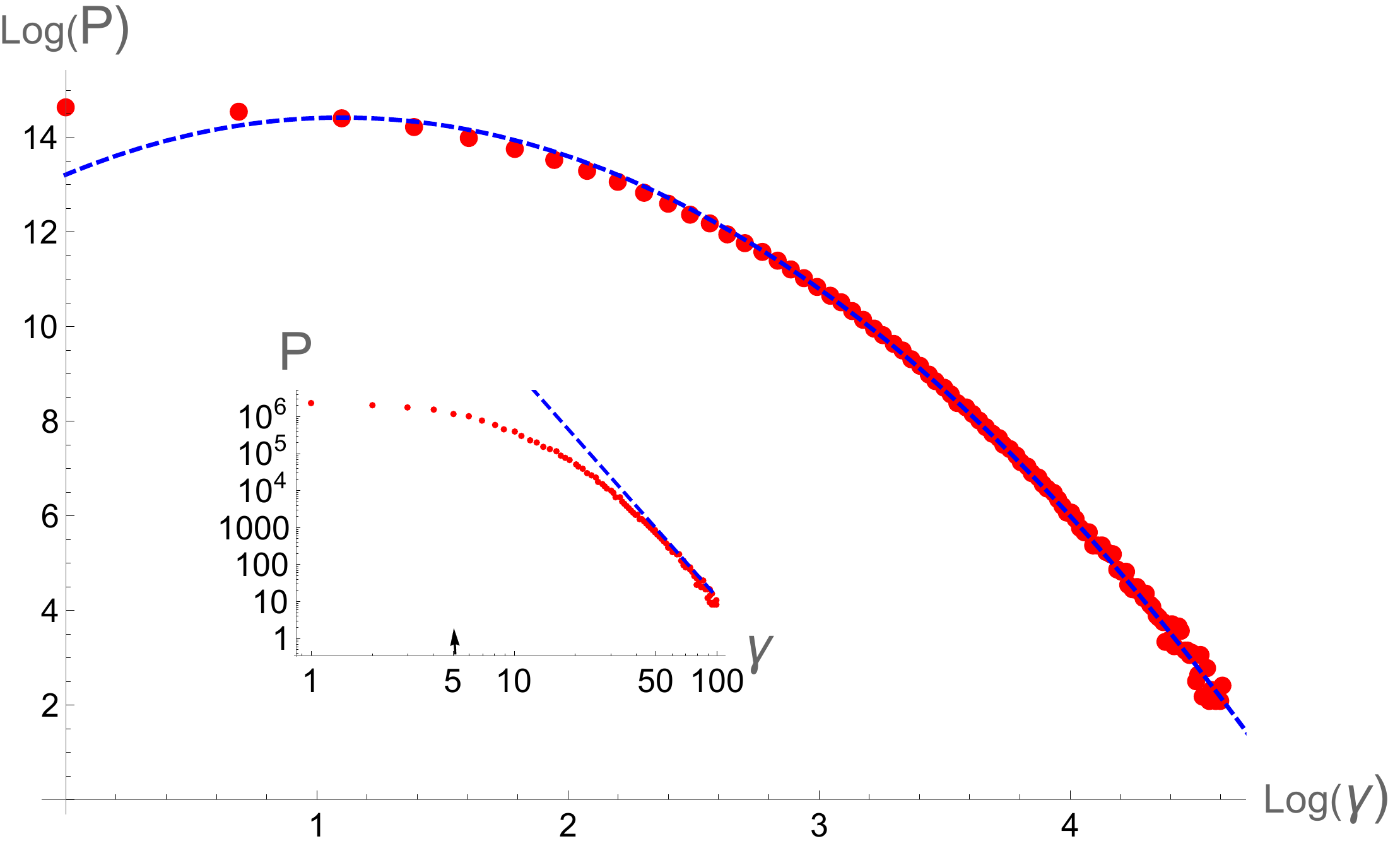}
  }
\caption{ Probability distribution $P(\gamma)$ of the acceleration $\gamma=dv/dt$ from Modane's data, in the domain $\gamma > 0$. (a) linear scale. (b) $\log$-$\log$ scale, the inset is to show  the fit with a power-law  $P(\gamma)= \gamma^{-6.2}$ (blue dashed curve), the main curve  ($\log(P)$ versus $\log(\gamma)$) displays a very good fit with the function $P(\gamma)/P_{0}= \exp[-(\log \gamma -1)^{2}]$ (blue dashed curve) on several decades. The standard deviation, indicated by an arrow, is $\sigma_{\gamma}=5.2$cm/$s^{2}$. Ordinate in a.u.,  abscissa $\gamma$ in cm/$s^{2}$.}
\label{fig:proba-gamma}
\end{figure}

As written above, the presence of singularities in turbulent flows could explain the non-cubic behavior of $\Psi_{2}(\tau)$ at the origin, and could also have a signature in the long tails of the probability distribution of the  acceleration $\gamma=dv/dt)$. We observed such long tails in the probability $P(\gamma)$, see  Fig.\ref{fig:proba-gamma} for positive $\gamma$ values ($P(\gamma)$ is an even function). 
 If one assumes that the velocity flow obeys a self-similar solution of the usual form with a single exponent (close to a singularity), we have shown in section \ref{sec:sing-accel} that the asymptotic behavior of $P(\gamma)$ at large $\gamma$ should be as $\gamma^{-2}$ . But the data are actually very well fitted by an exponential function of the  form $P(\gamma)=P_{0}  \exp[-(\log \gamma/\gamma_{0} )^{2}]$.   If one try to fit the data with a power law, it gives $P(\gamma)=\gamma^{-6.2}$ for large values of the acceleration (much greater than the standard deviation, $\sigma_{\gamma}=5.2$cm/$s^{2}$), see the inset. That  exponent is not the one predicted above by relation (\ref{eq:gamma-exp}) with $\alpha=1/2$, although $\alpha=1/2$ agrees with the $\Psi_{2}(\tau)$ behavior at small time.  We tried above to give an explanation for this apparent contradiction, see subsection \ref{sec:sing-accel},  by  attributing these two different behaviors to a  possible  singular solution of Euler equation  having a more complex dependence with respect to time than the one of usual self-similar solutions. Because of the present lack of understanding of the existence and nature of singular solutions of Euler equation of finite initial energy there is room for speculations on this subject, a freedom we use here. 

Let us now discuss another point in connection with $K41$ scalings. The successive increase and decrease of $\Psi_{2}(\tau)$ in Fig.\ref{fig:test-mod-irrev} are a priori surprising if one applies  standard Kolmogorov scalings  to the  test function. Actually the Kolmogorov description which leads to  power-law behavior (versus time and energy) of all quantities,  could not lead to a non monotonic behavior such as the one of $\Psi_{2}(\tau)$, on the contrary it predicts a power law behavior on a very large time interval  (including all the scales of the inertial range). To be more precise, let us derive how a function like $\psi_{2}(\tau)$ should behave if one assume the  Kolmogorov scalings. That should imply that any averaged quantity depending on a time difference $\tau$ can be written as,
\begin{eqnarray}
\label{e1}
 \Psi(\tau) \sim \tau^{\alpha} \epsilon ^{\beta},
\end{eqnarray}
a power function of $\tau$ times a power of $\epsilon$, the energy dissipated per unit mass and time in the turbulent flow, with the physical dimension $[\epsilon ] = L^2 T^{-3}$, 
$L$ dimension of length and $T$ dimension of time. 

The scaling laws  are different according if 
Taylor hypothesis of frozen turbulence is valid or not. Frozen turbulence requires that  velocity fluctuations have an intensity much smaller than the average velocity of convection in the mean flow. With Taylor hypothesis, time dependent records are images of space dependent quantities (variable $y$)  with the correspondence $\tau =y/v_{0}$ where $v_{0}$ is the convection velocity. Therefore the scaling laws must be derived by taking the distance $y$ along the flow as a dimensionalizing quantity. This yields the power law 
\begin{eqnarray}
\label{e2}
  \Psi(\tau)_{Taylor} \approx \epsilon y   \approx  v_{0} \epsilon   \tau  \textrm{.}
\end{eqnarray}
 This power law  
 is obviously hard to reconcile with the observed function which displays an extremum.  Note that in the case of Modane's data, we did observe such linear behavior, but only close to the origin. Moreover, as noted in the previous subsection, the linear (non cubic) behavior of $\Psi_{2}(\tau)$ close to $\tau=0$ implies a divergence of the acceleration. 
 Relying this feature on possible finite time singularities of solutions of the fluid equations, we infer  that the linear
 dependence of $\Psi_{2}$ for small $\tau$ could be attributed to the occurence of  singularities in the turbulent fluid.  
 
If one drops Taylor hypothesis and assumes that the scaling is with the time instead of  the distance, one finds
\begin{eqnarray}
\label{e3}
 \Psi(\tau) \approx \epsilon^{3/2} \tau^{3/2}  \textrm{.}
 \end{eqnarray}

In summary the  $K41$ Kolmogorov scalings which are supposed to describe the whole inertial range are in contradiction with the fact that any correlation has a finite range.  Nevertheless it is worth pointing out that the velocity fluctuations in Modane's wind tunnel have been used \cite{modane} to obtain one of the most accurate measurement of the energy spectrum of turbulent fluctuations  fitting very well the Kolmogorov-Obukhov scaling laws.

 This difficulty of reconciling the data with simple scaling laws goes beyond the  test function $ \Psi_{2}(\tau)$.  Let us consider for example the second order auto-correlation of the velocity fluctuations, $\Gamma_{v}(\tau)= <(v(t) - v_{0})(v(t + \tau) - v_{0})>$ which we did measure from the experimental data  (see Fig. \ref{fig:correl}).  Using the same scaling arguments as above, one finds that with Taylor hypothesis (the fluctuation $v(y, t)$ should be practically independent on $t$) this correlation (in space with $ y = v_{0}  \tau $) should scale like 
 \begin{eqnarray}
\label{e4}
< (v(t, y) - v_{0})(v(t, y +v_{0} \tau) - v_{0}) >_{{Taylor}} \approx(v_{0} \epsilon \tau)^{2/3}
\textrm{,}
 \end{eqnarray}

whereas dropping  Taylor hypothesis, the correlation should scale as
 \begin{eqnarray}
\label{e4}
< (v(t, y) - v_{0})(v(t, y+v_{0} \tau) - v_{0}) > \approx \epsilon \tau
\textrm{.}
 \end{eqnarray}
None of those scalings can be reconciled with any correlation function which has a maximum at $\tau=0$, and limited time range, much smaller than the full range  of the data.   
As written above, in the case of Modane's data, $C(\tau)$  displays a central peak decaying monotonously from a finite value at $\tau = 0$ to zero at  time of order $50ms$, larger than the short time scale associated to the Kolmogorov length where viscosity becomes significant, and much shorter than the large time scale, namely the  turnover time of the flow across the wind tunnel.

Let us point out another difficulty. Actually  the existence of non vanishing function  $\Psi_{2}(\tau)$ is hard to explain within Taylor hypothesis, because these oscillations are oscillations in space (not in time) with an asymmetry due to irreversibility. In one hand spatial asymmetry is consistent with the well known fact that skewness, the third moment of the single point velocity fluctuation, $\delta v(y,t)=v(y,t) -v_{0}$ is not zero. The relation $<\delta v(y,t)^{3}> \ne 0$ implies that, by some process the mean flow is correlated with the velocity fluctuations, meaning that this mean flow does more than just advecting the turbulent fluctuations by a Galilean transform. Said otherwise asymmetry indicates that there is a memory in the turbulent fluctuations of the way they are generated, namely of  the direction of the velocity with respect to the walls whose interaction with this flow generates turbulence. But on the other hand  the existence of a non zero skewness goes against Taylor hypothesis: if turbulence is only convected by the mean flow, there is no reason for a correlation between the direction of the fluctuations and this mean flow.

\section {Test for  turbulent energy cascade}
\label{sec:cascade}
In this section we use well chosen self-correlation functions to show that the transfer of energy goes from large scale to small scales, as expected from Kolmogorov theory, with a delay time increasing as the intermodal distance increases in the wave-number space (distance between the donor mode and the receptor mode).  

 In the framework of Taylor hypothesis, the density of kinetic energy in wave number space, $E(k,t_{0})$,  in the interval $[k, k+{\mathrm{d}}k]$ at time $t_{0}$, can be derived from the velocity $v(t)$ at a single point. It is proportional to the square modulus of the Fourier transform of $v(t)w(t-t_{0})$, where $w(t)$ is a window function centered around $t_{0}$ of appropriate width. Using relation (\ref{eq:taylor}), it gives
 \begin{equation}
E(k, t_{0})= \vert  \int \mathrm{d}t  \ \ v(t) w(t-t_{0}) e^{ikv_{0}t} \vert ^{2}
 \mathrm{.}
 \label{eq:Ek}
\end{equation}
We shall use a Gaussian window which must include many oscillations of wavenumber $k$, 
 \begin{equation}
  w(t) =  \frac{1}{\sqrt{2\pi \theta}} e^{- \frac{t^2}{\theta^2}}
   \mathrm{.}
 \label{eq:window}
\end{equation}
Practically the width $\theta$ has to be larger than $1/kv_{0}$,  and much smaller than the integration time.
As written in the introduction, one specificity of the cascade lies in the fact  that the  energy density $E(k_{1},t)$ takes some time to be transferred to  the frequency domain $k_{2}= k_{1}+\Delta k$ (with $\Delta k$ positive) and more time to reach the domain $k'_{2}=k_{1}+  \Delta k'$ where $\Delta k'  > \Delta k$. This dependence of the time delay with respect to the distance between the $k$'s should have a signature in the time dependent cross-correlations introduced in \cite{pom82}: 
\begin{equation}
\mathcal{C}(k_{1}, t_{1}; \; k_{2}, t_{2})= <E(k_{1},t_{1}) E(k_{2}, t_{2})> - <E(k_{1})> <E(k_{2})>
 \mathrm{,}
 \label{eq:corelk}
\end{equation}
 which will not be an even function of $t_{2}-t_{1}$. 
In the following we shall consider $k_{2}>k_{1}$ and $t_{2}> t_{1}$ with  $k= k_{2}-k_{1}, t =t_{2}-t_{1}$.   
Note that time dependent correlations must be considered cautiously to make a clear distinction between the oscillating character of exchanges of energy in the system, and the
 transfer of energy from one scale to the other in  the forward time direction. The latter effect can be probed by looking at the test functions 
 \begin{equation}
H (k_{1},k_{2}, t)= <E(k_{1},t_{1}) E(k_{2}, t_{2})>  - <E(k_{1},t_{2}) E(k_{2}, t_{1})> 
\mathrm{.}
\label{eq:testH}
\end{equation}
Focusing on a given domain of the spectrum, we  consider below the test function $H$ for  specific values $k_{1}$,  
\begin{equation}
H_{k_{1}}( k, t)= \mathcal{C}(k_{1},k_{2}, t)-\mathcal{C}(k_{1},k_{2}, -t)
\mathrm{,}
\label{eq:testHb}
\end{equation}
that restricts each calculation to  the two dimensional phase space $k,t$.
The function $H$ vanishes for $t=0$ and also
for large delays because $E(k_{2}, t_{2})$ becomes then statistically independent of $E(k_{1}, t_{1})$  and the correlation function (\ref{eq:corelk}) vanishes.  
In the regions where $H$ is positive, the mode $k_{2}$ is  more  strongly correlated to the mode $k_{1}$ for positive delay $t_{2}-t_{1}$  than for negative delay.

In order to prove that a turbulent cascade takes place, we have to check from the $H$ function that it takes more and more time  for the energy to go to more widely separated scales. The latter point is crucial because if, for example  the energy transfer from large to small scales is the result of a finite time singularity (like in the one dimensional Burgers equation), the time scale does not depend critically on the difference of space scales but on the time needed for the occurrence of a singularity, namely on the initial conditions, as written in the introduction.  However this claim is hard to reconcile with the possibility that the short time behavior of the auto-correlation function is a consequence of finite time singularities with a self-similar behavior. We plan to return to this crucial issue. 
Note that the self-correlation functions exactly cancel if the system is reversible and not otherwise. 
In the following a variant to the function $H$ will be our test function for proving the existence of turbulent cascade, it is given by the same expression as (\ref{eq:testH}) but with a function $E(k,t)= \vert  \int \mathrm{d}t  \ \ v(t) w(t-t_{0}) e^{ikv_{0}t} \vert $,  which is the square root of the energy density, that allows to reduce the noise of the signal.  

\subsection{Test on Modane experiment}

\begin{figure}
\centerline{
\includegraphics[height=2.5in]{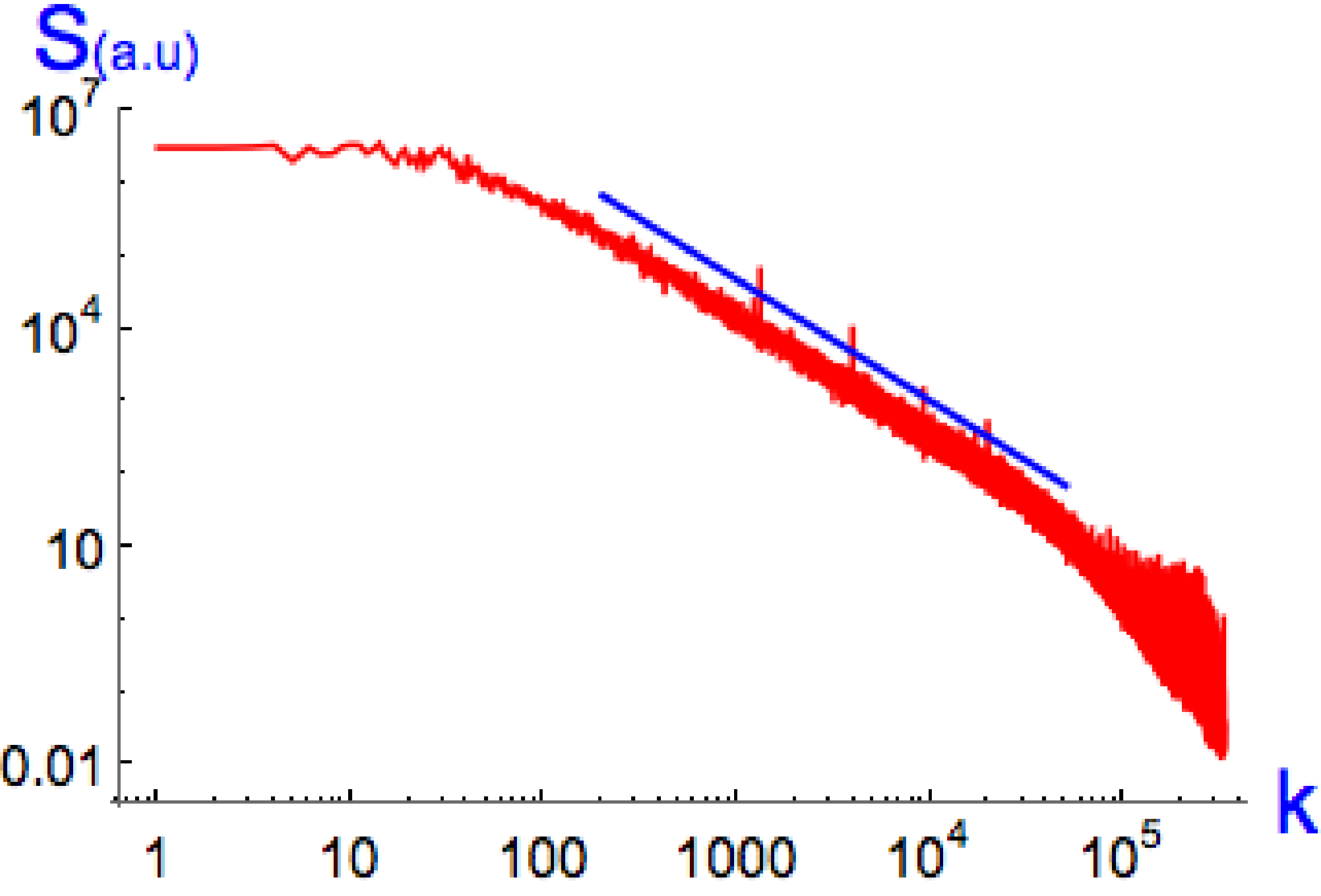}
  }
\caption{
Modane experimental spectrum $S(k)$ in arbitrary units, the abscissa $k$ is in units of $k_{u}=2\pi f_{u}/v_{0}$ where $v_{0}$ is the mean velocity of the flow, $f_{u}=10 n t_{s}$,  $n=2^{16}$ is the number of sampling times per file. It gives $k_{u}=0.011 m^{-1}$. The solid line  displays the $k^{-5/3}$ behavior of the spectrum in the inertial range. }
\label{fig:spectre}
\end{figure}

The spectrum  $S(k)$ is obtained by running average over successive  $t_{i}$ values of  Fourier transforms, $S(k)= \sum_{i}S_{i}$, where
 \begin{equation}
S_{i}=  \vert  \int _{T}\mathrm{d}t v(t-t_{i})  e^{ikv_{0}t} \vert ^{2}
\mathrm{,}
\label{eq:sp}
\end{equation}
with an the integration time   $T=26s$. It is
 shown in fig.(\ref{fig:spectre}) which displays an inertial domain extending over several decades. We estimate its large-k limit to be  about $k=50000$ in units of the figure (see caption). 

Let us now describe the results concerning the test function $H$.   To calculate  $H$ we first select  a wave-number value $k_{1}$ belonging  either to the inertial range, or to the dissipative one. The correlation functions   (\ref{eq:corelk})  are also calculated  by running  average over time $t_{1}$.  The width $\theta$ of the window function, equation (\ref{eq:window}),  has to be chosen with caution. In order to focus the analysis on a given band of  modes (of spatial frequency $k_{1}\le k \le k_{2}$), we take  a window  function five times larger than the inverse of the frequency $f_{1}= k_{1}v_{0}/2\pi$.  All three dimensional plots shown below display oscillations, with positive and negative amplitudes.  In the phase-space $k= k_{2}-k_{1}, t=t_{2}-t_{1}$, we interpret positive amplitudes domains as  regions where the correlation between frequencies  $k_{1}$ and $k_{2}$ is  stronger for  positive time than for negative time, namely regions where the transfer of energy towards small scales (large $k$ values) occurs for positive time, a process named \textit{direct transfer} in the following. Likewise, by  symmetry, negative amplitudes domains are seen as  regions where the correlation between frequencies  $k_{1}$ and $k_{2}$ is  stronger in the past than in the future, namely regions where the transfer of energy goes backward, i.e. towards large scales (small $k$ values), named below \textit{inverse transfer}.

\subsubsection{In the inertial range}

 \begin{figure}
\centerline{
(a)\includegraphics[height=2.50in]{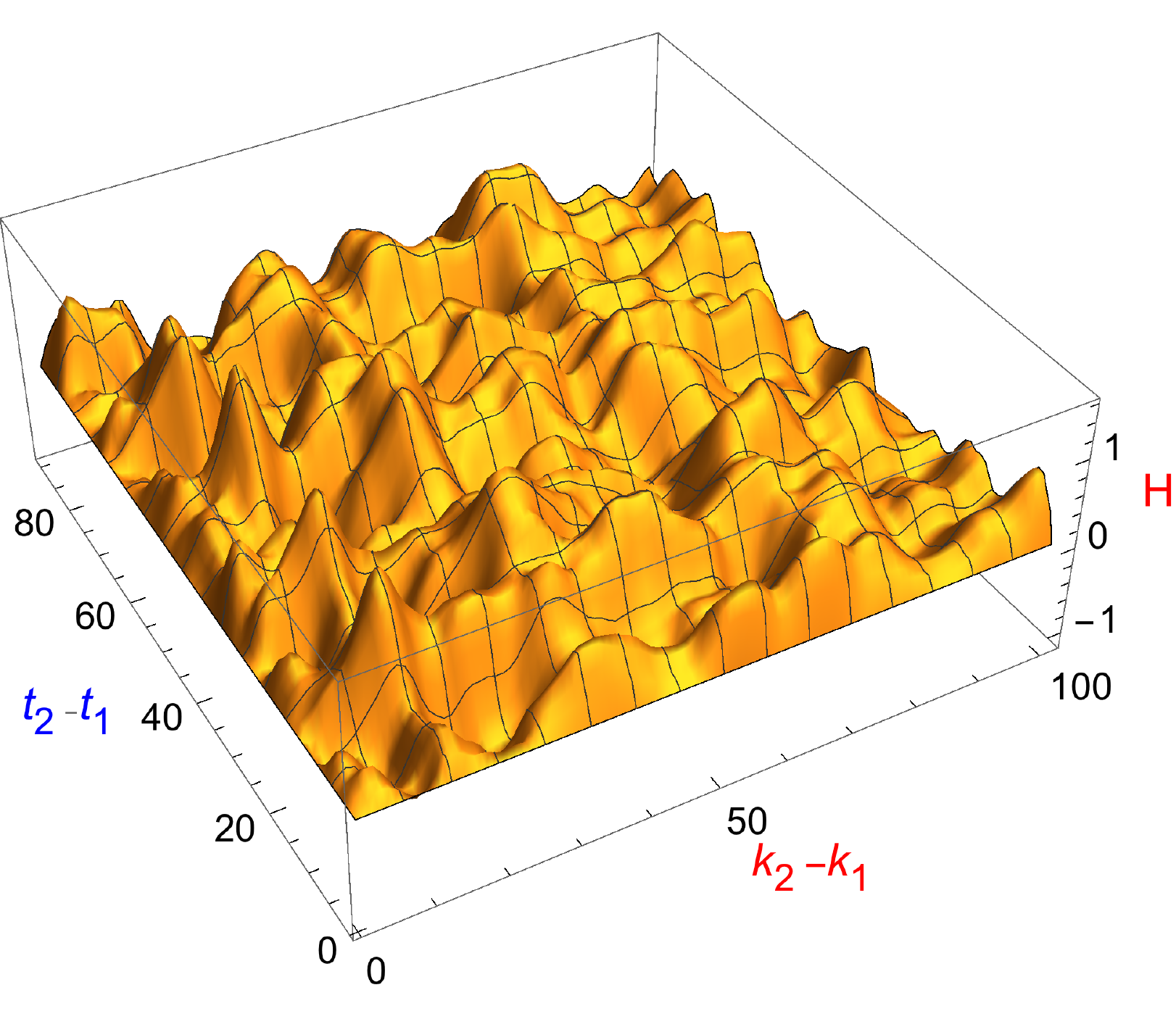}
}
\centerline{
(b)\includegraphics[height=2in]{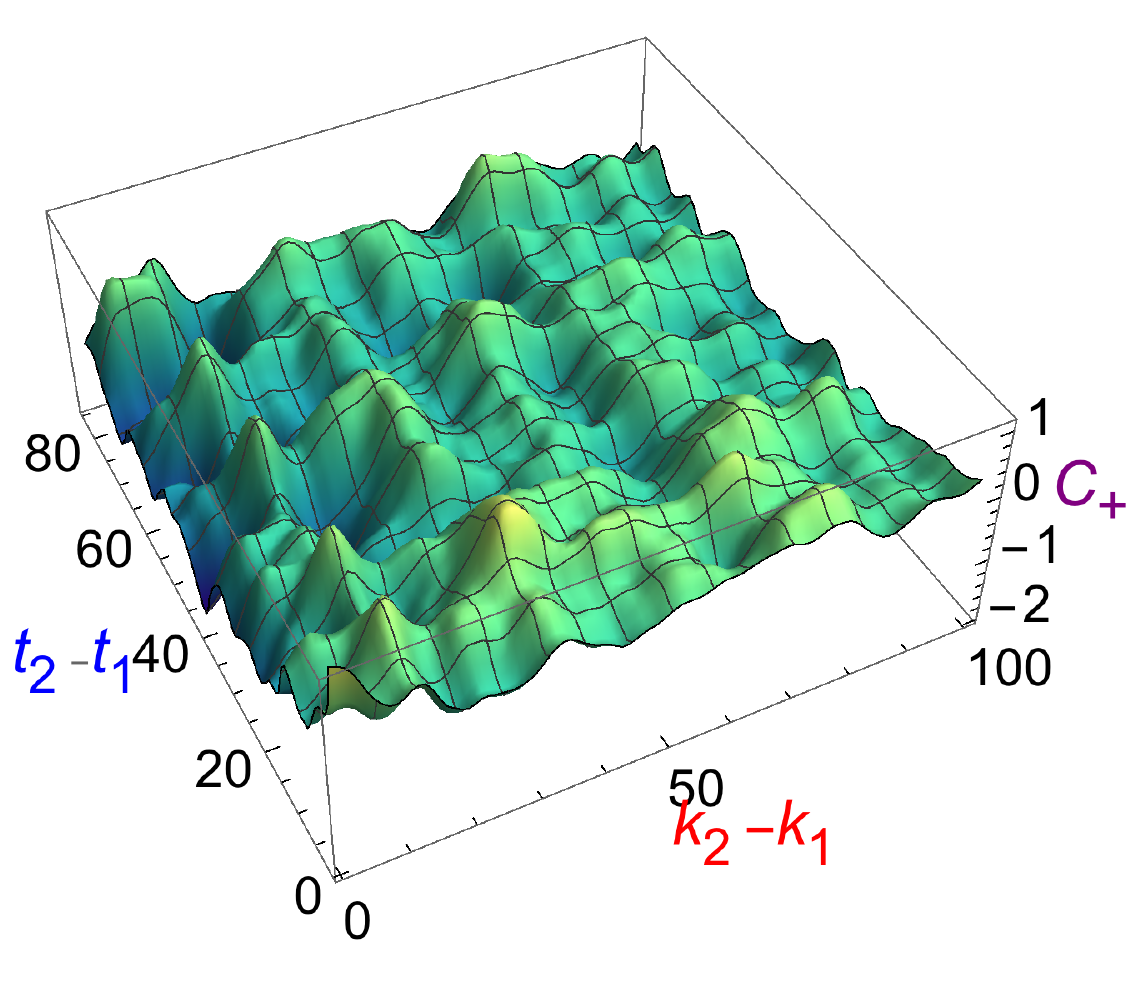}
(c)\includegraphics[height=2in]{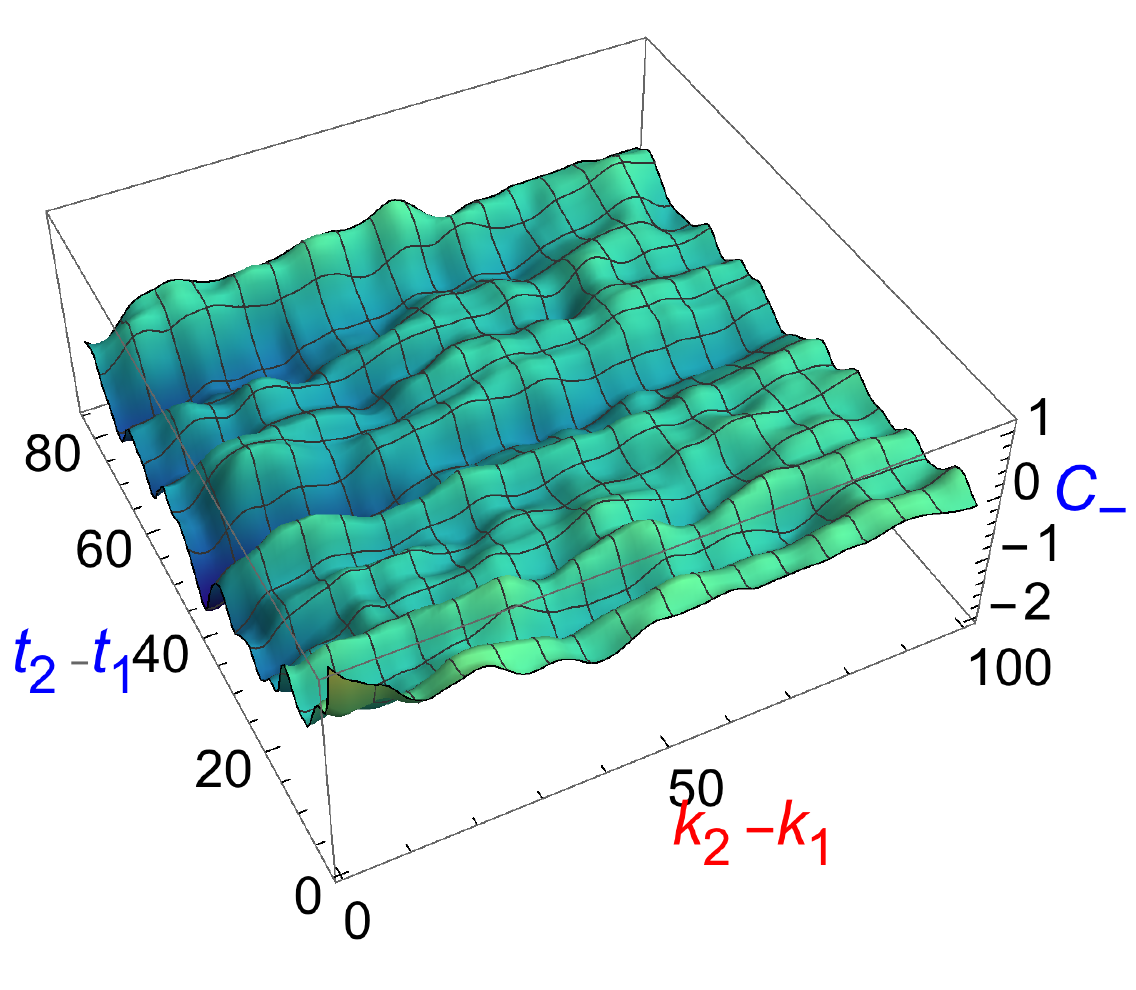}
  }
\caption{
Test  function $H_{k_{1}} (k, t)$  in (a) and k-correlation functions $C_{\pm}=  \mathcal{C}(k_{1},k_{2}, \pm t)$  in (b)-(c), calculated from Modane data, for $k_{1},k_{2}$ in the inertial range ($k_{1}=6182$ in units of Fig.\ref{fig:spectre}). In units of the present figure  $k_{1}=50$,  then $k_{2}$ extends up to $3 k_{1}$. The time axis is in units of $50t_{s}$ (real time  in $ms$ is $t_{ms}= 2(t-1)$). $H$ is in arbitrary units in all plots. In (b)-(c)  the region $t < 7$ close to the origine  is  not shown for clarity. 
}
\label{fig:testH-K50}
\end{figure}

Fig.\ref{fig:testH-K50} shows the three dimensional plots of the three functions $H$, $\mathcal{C}(k_{1},k_{2}, \pm t)$ for  wave-numbers $k_{1},k_{2}$ belonging to the inertial domain, see caption. Here the range of variables $k,t$ is large.   The $k$-axis  extends up to $3k_{1}$ in order to capture possible correlations of  the energy density at  the wave-number $k_{1}$ with the energy density of the second, and third harmonic of $k_{1}$. The time axis extends up to $180ms$ which is equal to $4 \tau_{c}$.  The two  functions $C_{\pm}= \mathcal{C}(k_{1},k_{2}, \pm t)$  used to define  the test function $H$, are shown in order to put into evidence their very different behavior: $C_{+}$ and  $C_{-}$ display bumps for a quite well defined set  of  $k_{i}$ values, but in  $C_{+}$ the bumps have specific dynamics, although time oscillations  in $C_{-}$  concern the whole quasi-frozen  ensemble of waves. We observe that the amplitude of the oscillations are of same order in these two plots, it follows that their difference $H$ is also of same order, an important point which shows that the test function displays a significant signal. Let us precise that the small time region is suppressed in the  three dimensional plots (b) and (c) to make the oscillations visible, that was not done in (a) because the test function $H$ is null at time $t=0$ by definition, contrary to the functions $C_{\pm}$ which have a  maximum at the origin.

\begin{figure}
\centerline{ 
(a)\includegraphics[height=1.5in]{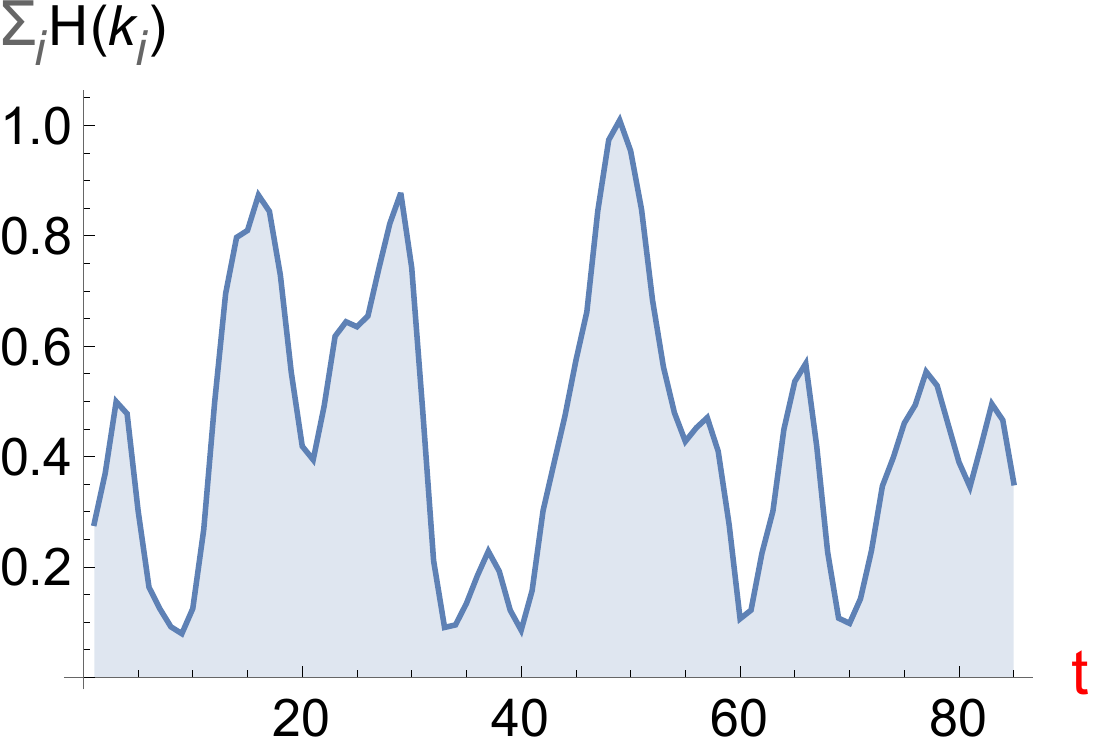} 
 (b) \includegraphics[height=1.5in]{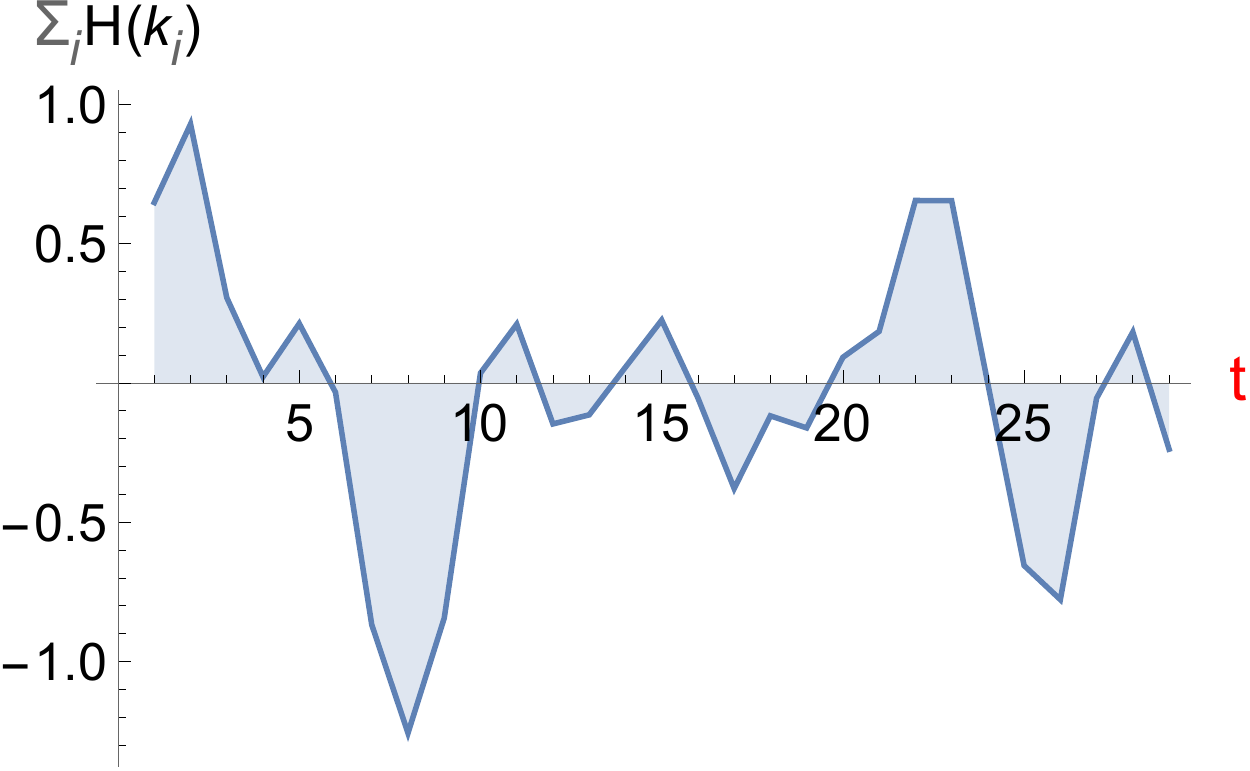}
 }
\caption{
$k$-integral function $\mathcal{H}(t)$  (a) corresponding to Fig.\ref{fig:testH-K50},  for modes in the inertial range; (b) corresponding to Fig.\ref{fig:testH-limit}, for modes at the frontier inertial/dissipative range.  $t$ in same units as in Fig.\ref{fig:testH-K50}.
}
\label{fig:kinteg}
\end{figure}

 Another important remark is that $H$ is positive on a large part of the $(k,t)$ domain, a property in favor of a transfer of the energy from $k_{1}$ to smaller scales, as announced above.  As time evolves, a detailed analysis of Fig.\ref{fig:testH-K50}(a) shows that the function $H$ becomes  transiently (and almost periodically) negative, that shows an inverse  transfer for these modes (around $k=20$, not clear in Fig.(a)).  These phases of inverse transfer are always followed by a strong increase of $H$ in other k-regions. This back-and-forth motion is the matter of our study. 
 In order to visualize when this occurs, we introduce the integral  of $H$ over $k$ (restricted to the $k$-domain of our study), a new function of time called $\mathcal{H}(t)$,
\begin{equation}
\mathcal{H}(t)= \sum_{i} H(k_{i},t)
\mathrm{.}
\label{eq:sumtHk}
\end{equation}
As time evolves, this function is expected to decrease whenever $H$ get negative amplitude domains, and to increase when these negative amplitude domains disappear. We also infer that the maxima of $\mathcal{H}$ occur when $H$ has modes  strongly correlated with $k_{1}$,  namely when there is a transfer  of energy (from the mode  $k_{1}$  to  one(s) with larger $k$ value (direct transfer). This function $\mathcal{H}(t)$, named below $k$-integral function,  is drawn in Fig.\ref{fig:kinteg}(a). It displays  successive maxima and minima, as expected. We  note that $\mathcal{H}(t)$ is positive, which is in favor of global direct transfer.

\begin{figure}
\centerline{ 
(a)\includegraphics[height=1.50in]{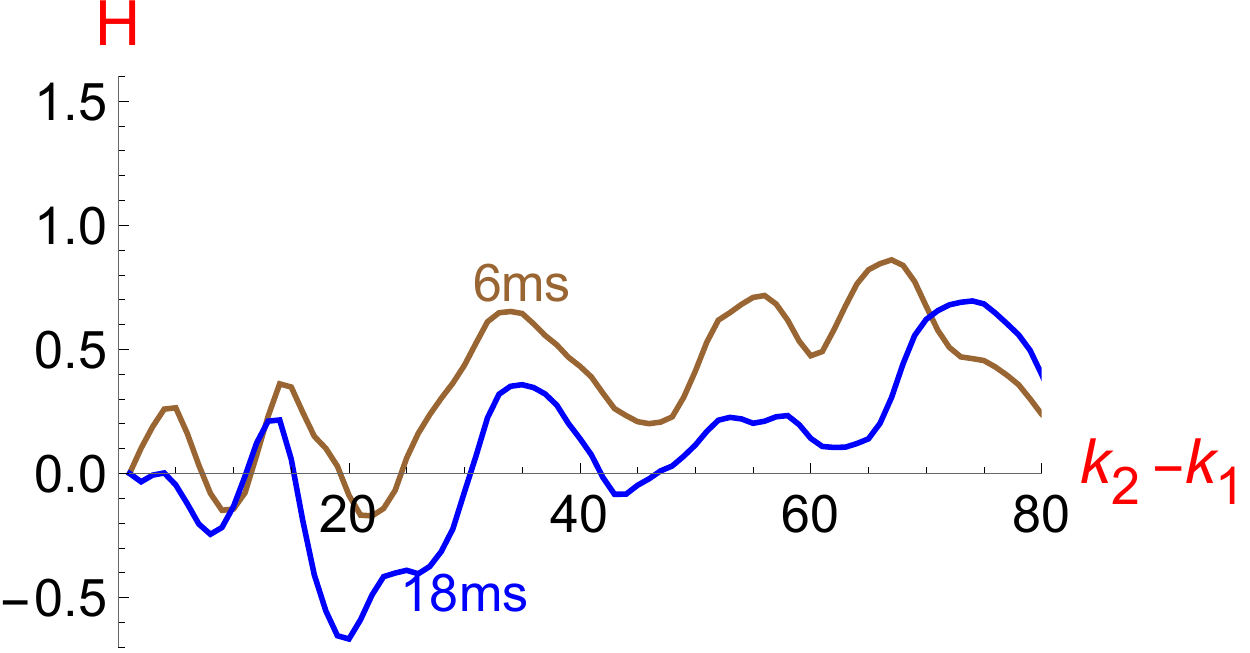}
(b)\includegraphics[height=1.50in]{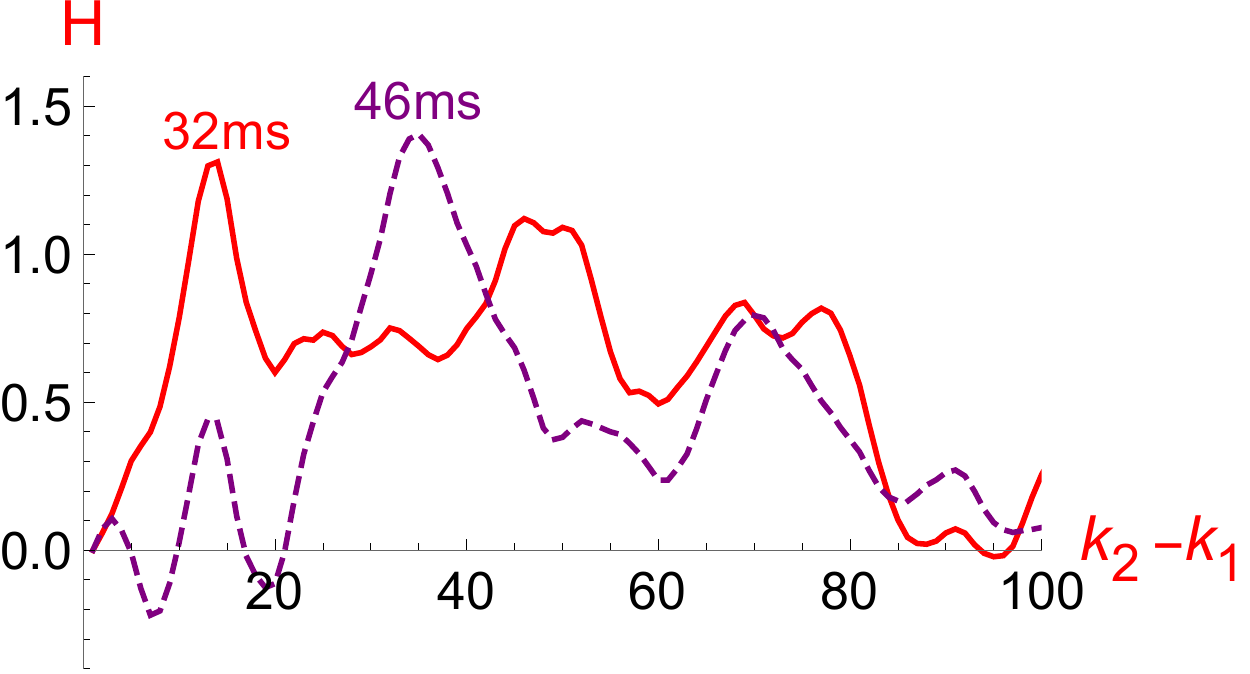}
  }
\caption{
k-profiles of the test  function $\mathcal{H}$, (a)   at early times $t=4$ and $t=10$ in units of Figs.\ref{fig:kinteg} , (b) at times $t=17$ and $t=24$ corresponding to the emergence of the first and second maximum of $H(k,t)$.  $k,t$ in same units as in Fig.\ref{fig:testH-K50}.
}
\label{fig:prof-50}
\end{figure}

\begin{figure}
\centerline{ 
(a)\includegraphics[height=1.0in]{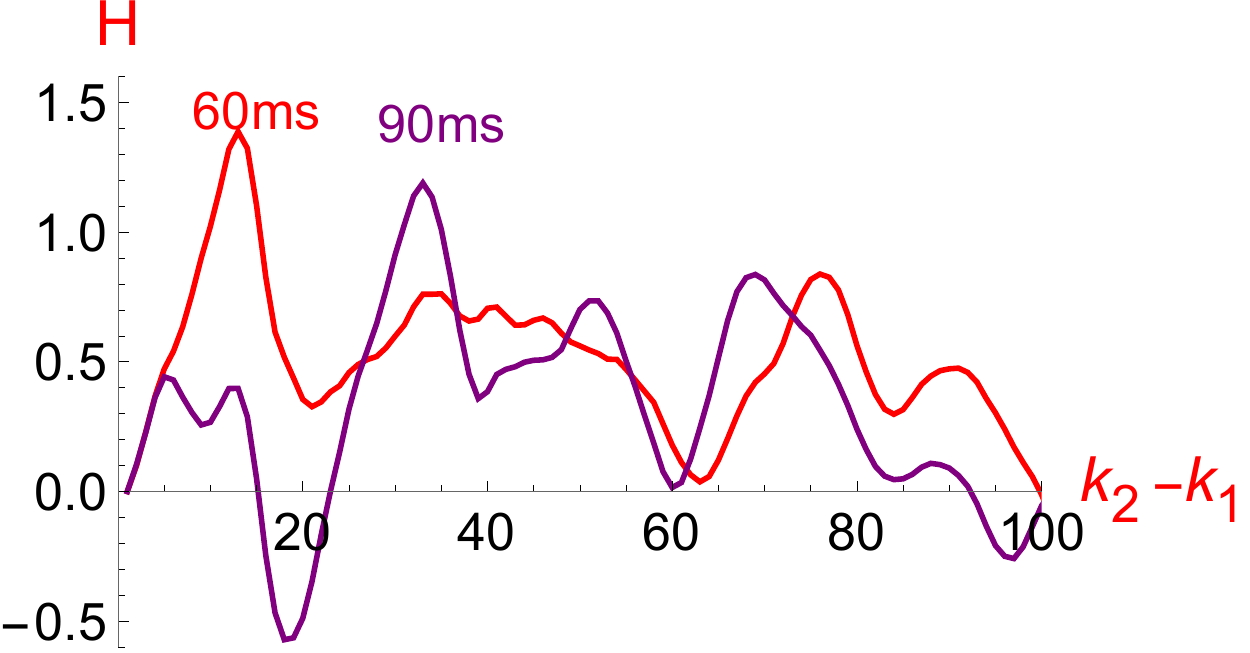}
(b)\includegraphics[height=1.0in]{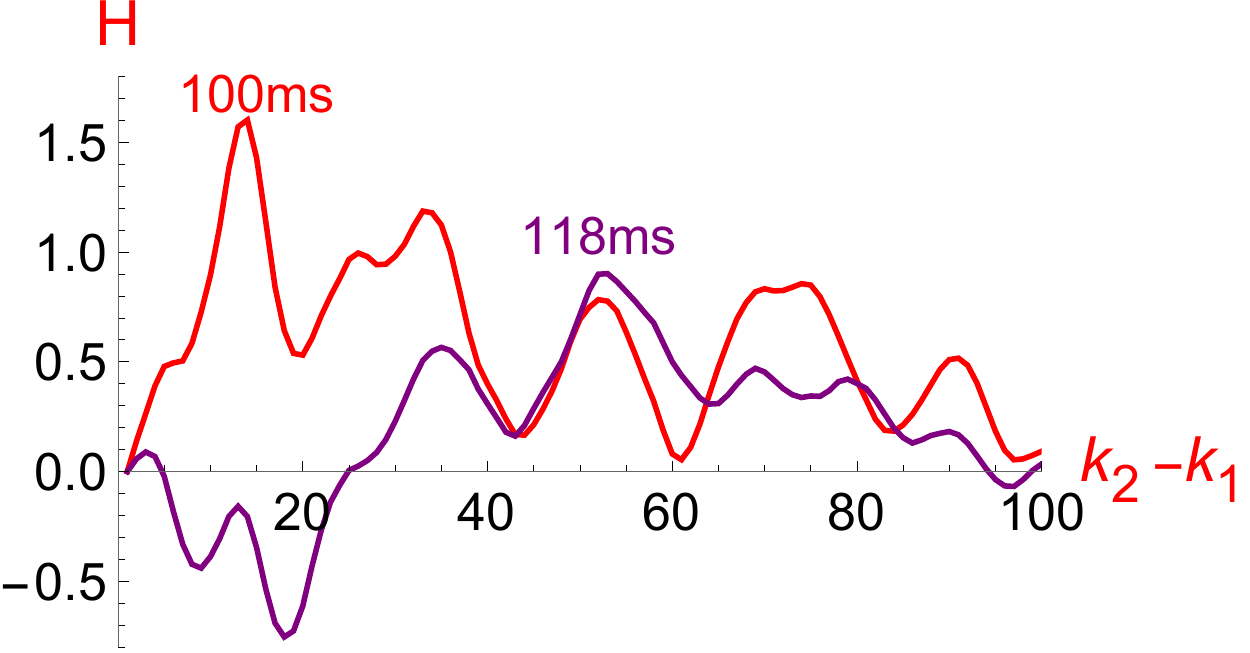}
(c)\includegraphics[height=1.0in]{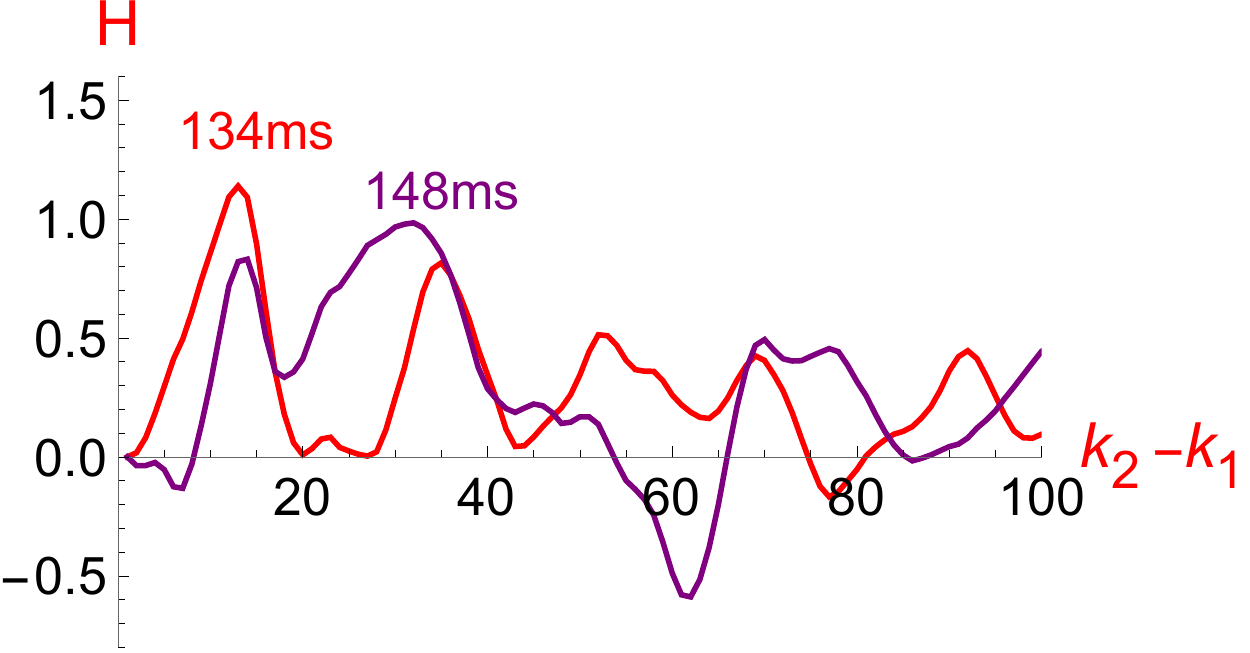}
  }
\caption{ 
k-profiles of the test  function $\mathcal{H}$, at transfer times following the ones of Fig.\ref{fig:prof-50}, (a) for $t=31$ and $t+46$, (b) for $t=52$ and $t=60$, (d) for $t=68$ and $t=75$, in units of Figs.\ref{fig:kinteg}corresponding to the emergence of the successive maxima of $H(k,t)$.  $k,t$ in same units as in Fig.\ref{fig:testH-K50}.
}
\label{fig:prof-50s}
\end{figure}

As written above, the evidence of direct transfer (to small scales) is not enough to claim that Kolmogorov cascade exists. That should be stated if, in addition, one can prove that a first transfer from $k_{1}$ to  $k_{2}$ is followed by a second transfer from $k_{1}$ to other modes with $k'_{2} > k_{2}$. The signature of such an event could be find
by looking at the  $k$-profiles of $H(k,t_{i})$  at successive times $t_{i}$, because the maxima of $H$ (in the space $k,t$ ) correspond to maxima of  energy transfer, as already written.
We have therefore studied the $k$ profiles of $H$,  and reported some relevant ones 
 in Fig.\ref{fig:prof-50}. Fig.\ref{fig:prof-50} (b) is the main result of this subsection. It shows the k-profiles  when $H$ reaches its first and second maximum.  As indicated in the figure, the two first maxima of $H$ are shifted in time,  and concern different bands of frequencies: the first transfer is mostly towards the modes around $k_{2}=k_{1}+15$, the second one is to $k_{2}=k_{1}+35$. Therefore this figure shows  that the transfer of energy from $k_{1}$ to the band of modes around $k_{1}+ 15$ (plus other maxima at larger frequencies, see another example below), takes place before the transfer from $k_{1}$ to  towards  $k_{1}+35$. 
 Note that the maxima of $H$  and $\mathcal{H}(t)$ occur approximately at  the same times ($t=17$ and $24$ in units of Fig.\ref{fig:kinteg}), that justifies to use the $k$-integral function for a rapid investigation of the time and direction of the transfers. The profiles in Fig.\ref{fig:prof-50}(b) show that other modes  receive energy from $k_{1}$, but with less efficiency. For instance  at time $t=32$ms, the mode $k=50$ which is the first harmonic of $k_{1}$, and the modes around $k=70$ receive also energy from $k_{1}$.
In between these two direct transfers, the amplitude of $k$-profiles of $H$ decreases,  getting negative domains, but the sum over $k$ remains mostly positive. 
We have also analyzed the k-profiles at later times. They display new  direct transfers (from $k_{1}$) towards  larger $k$ values, as illustrated in Figs.\ref{fig:prof-50s}. The transfers also occur when $\mathcal{H}(t)$ is maximum. Note that these late $k$-profiles have also a small negative amplitude region, suggesting a more complex process with co-existence of forward and backward  (in $k$-space) exchange of the energy. 

Let us also consider the history before the first direct transfers.
 Fig.\ref{fig:prof-50}(a) shows the evolution of the k-profile of $H$ at early times. 
At small times the amplitude of $H$, which is small and mostly positive,  grows  in a large $k$-domain until $t=4$ ($6$ms), then it decreases and a negative $k$-domain appears around $k=20-30$, getting a maximum surface at time $t=10$ ($18$ms) which is  the abscissa of the first minimum of $\mathcal{H}(t)$, as expected. In other words this early stage displays a (small) direct-then-inverse transfer of energy.  After this tide-like motion (rising tide followed by ebb tide), the negative domain disappears, $H$  becomes positive elsewhere, with well-separated maxima at different times. 
  
  In summary the direct cascade of energy does exist for the modes belonging to the inertial domain, this occurring after a sort of tidal motion of small amplitude where the rising tide is followed by an ebb tide, then a rising tide of large amplitude starts, the one  which leads  to the two time-shifted transfers shown in Fig.\ref{fig:prof-50}(b).

\begin{figure}
\centerline{ 
(a)\includegraphics[height=1.5in]{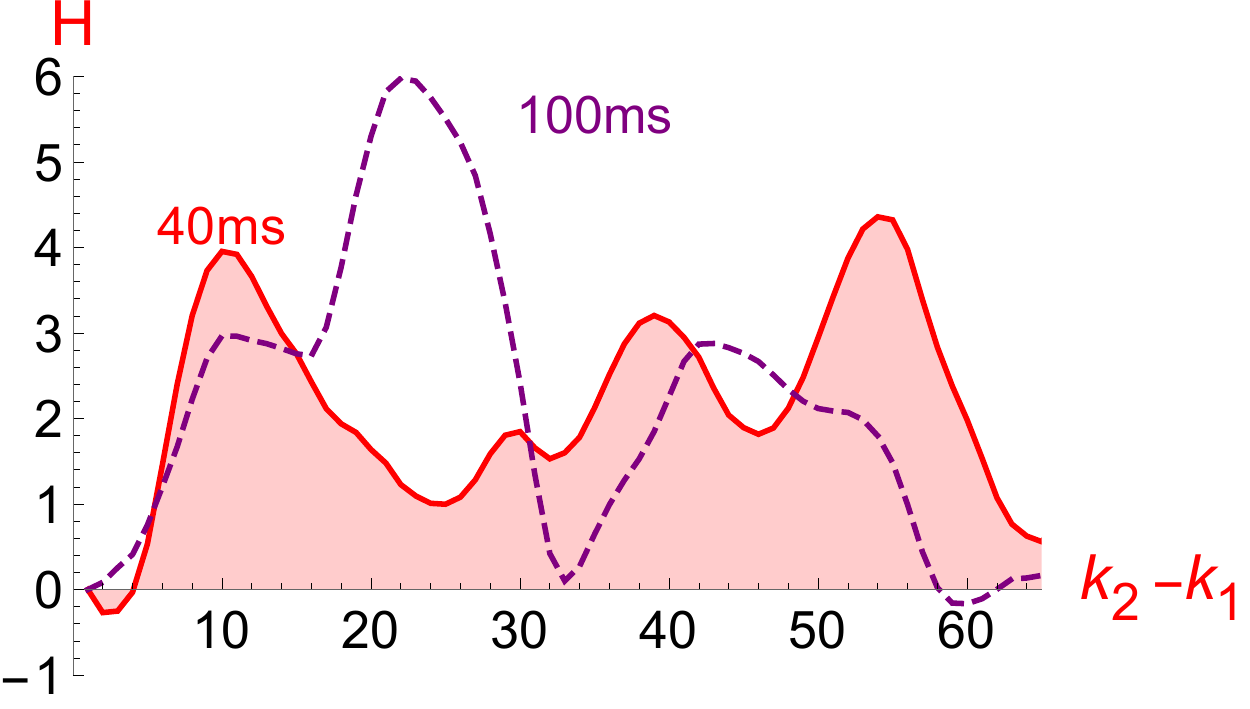}  
(b)\includegraphics[height=1.5in]{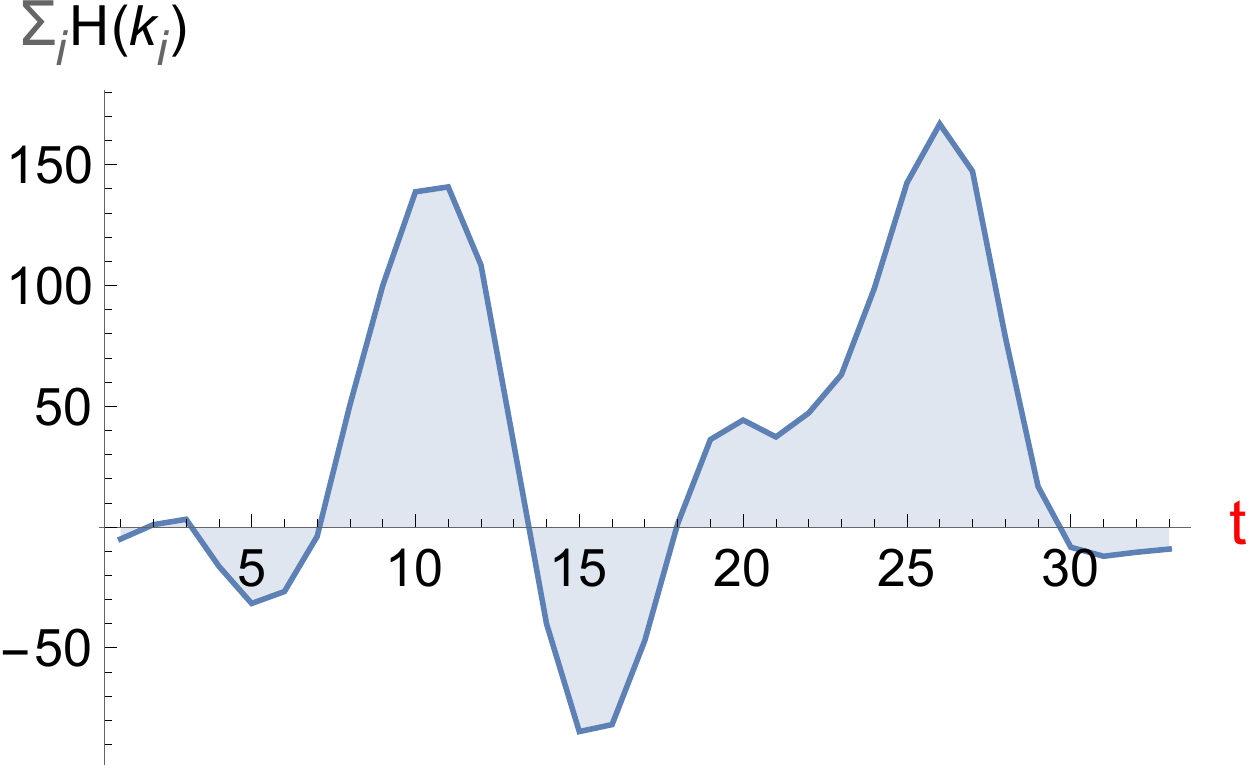}
  }
\caption{ Multi-band direct transfer in the inertial range.  $k_{1}=3100$ in units of Fig.\ref{fig:spectre}  ($k_{1}=50$ in units of the figure). Real time is  $\tau_{ms}=2(t-1)$.
(a) k-profiles of the test  function $\mathcal{H}$ at  times corresponding to the emergence of successive (in time) maxima of $H(k,t)$. (b) $k$-integral function showing the average time delay between transfers $\tau'=50$ms.
}
\label{fig:no-casc-inert}
\end{figure}

To complete this claim we have to note that  besides this clear cascading transfer towards successive smaller scales, we have observed several cases of multi-band transfers, involving more than a single acceptor mode. An example  is shown in Fig.\ref{fig:no-casc-inert} where the test function $H$ has two maxima with quasi-equal amplitude at first transfer.  
In the present case, and in several other cases that we have investigated  (see also Fig.\ref{fig:prof-50}), there is a small $k$ mode involved in the first transfer together with a large band of modes around the second harmonic ( $k =50$), that could illustrate the naive picture of a structure dividing into two parts (of similar dimensions). From Fig.\ref{fig:no-casc-inert} one can assert that  the transfer from $k_{1}$ to $k_{1}+10$ occurs before the transfer from $k_{1}$ to $k_{1} +25$,  which is a true cascading transfer, but we have to notice that  the second harmonic of $k_{1}$ is another acceptor mode. In summary   
this result does not contradict the previous exemple of clear cascading process,  but it
 shows the complexity of the process of transfer of energy even when focusing on the inertial range. 
 
The $k$-integral shown in Fig.\ref{fig:no-casc-inert}(b), is positive almost everywhere and displays successive maxima, which are separated by an average time delay about $\tau'=50ms$. 
In the previous case where cascading process was observed, taking the time delay of the first direct transfer as characteristic of the transfer, we had $\tau= 32ms$ for  $k_{1}= 6182$. Here we have  $\tau '= 50ms$  for  $k'_{1} = 3091$ (same units).  The ratios
\begin{equation}
\frac{\tau}{\tau'}=1.57 \qquad   ( \frac{k_{1}}{k'_{1}})^{-2/3} =1.59
\mathrm{.}
\label{eq:sc}
\end{equation}
are in  good  agreement with the scaling laws (\ref{eq:taul}) predicted by $K41$.
  
This example illustrates the difficulty of such a numerical study  to answer the question whether the cascading transfer is fully operative in the whole inertial domain, or not. Besides the fact that  we have limited datas, we must notice that  the choice of the width of  the window function is  delicate, because a bad choice  could amplify or reduce the amplitude of some frequencies. We did make several tests to ensure the validity of our results. 

\subsubsection{Transition domain between the inertial and the dissipative range}
\begin{figure}
\centerline{
(a)\includegraphics[height=2.0in]{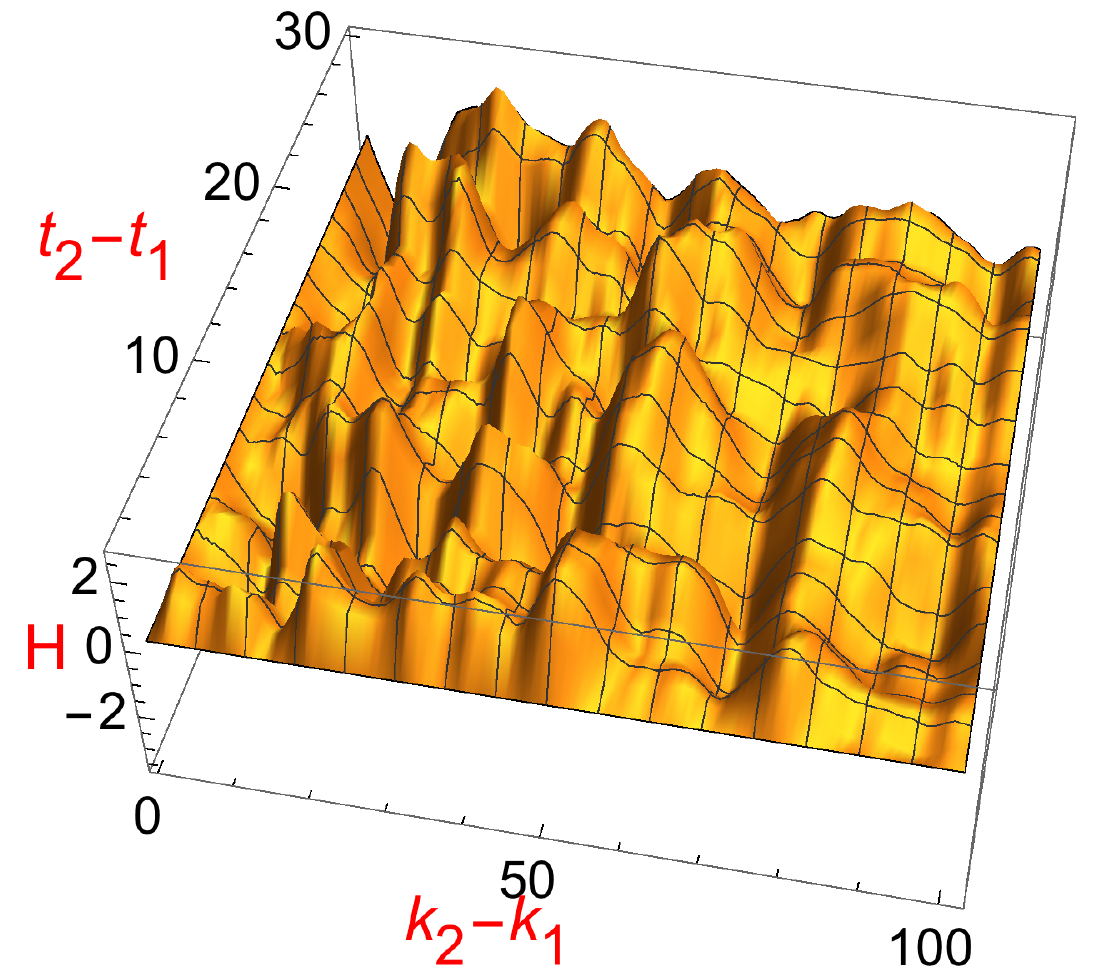}
}
  \centerline{
(b)\includegraphics[height=1.5in]{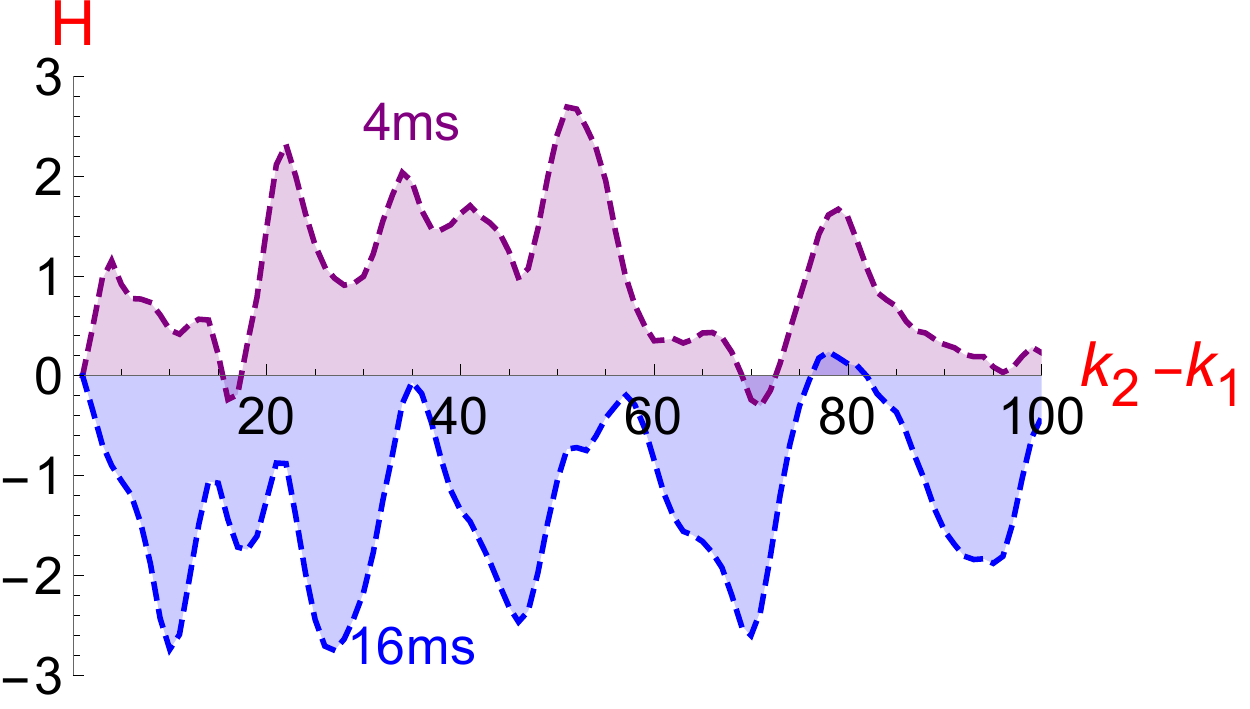}
(c)\includegraphics[height=1.5in]{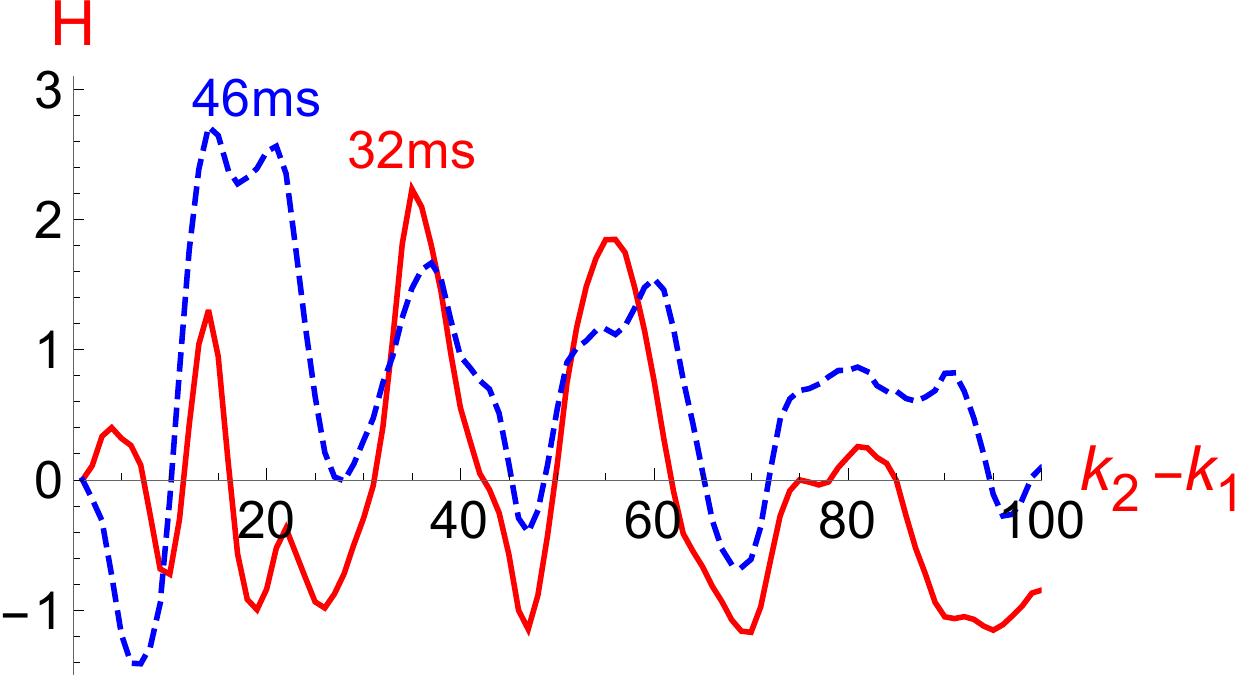}
  }

\caption{
Test  function $H (k_{1},k_{2}, t_{2}-t_{1})$  and $k$-profiles for $k_{1,2}$ values at the frontier between  the inertial and dissipative range, for $k_{1}=17250$ in units of Fig.\ref{fig:spectre}. (a) three dimensional plot; (b)-(c) $k$-profiles $H (k_{1}-k_{2}) $  at given times. Fig (b)  at $t=2$ and $t=8$ in units of (a) shows a direct transfer followed by an inverse one for the whole band of modes. This tide-like motion duration is $16$ms.  In Fig. (c)  the solid red line  displays a mixed transfer characterized by positive and negative extrema. The dashed  blue curve   corresponding to the second maximum of the $k$-integral function, $t=23$ in Fig.\ref{fig:kinteg}, shows that the inverse transfer is damped,  the direct one remains (from $k_{1}$ to separate  modes). Time is scaled to $2$ms $t_{ms}=2t$,  $k_{2}$ extends to $3 k_{1}$}.
\label{fig:testH-limit}
\end{figure}

We ask the question: does this cascading effect persist beyond the inertial range? Fig.\ref{fig:testH-limit}  shows the test function $H$ for  $k$-values located in the vicinity of the boundary between the inertial and the  dissipative domains.  In units of  Fig.\ref{fig:spectre}, we have $k_{1}= 17250$ and $k_{2}$  varies up to $k_{2}=3 k_{1}$,  considered above as the frontier of the dissipative range.  The time extension is $60$ ms.  The three dimensional plot in Fig. (a) displays oscillations which are of the same order of magnitude as each function $ \mathcal{C}(k_{1},k_{2}, t)$ and $ \mathcal{C}(k_{1},k_{2}, -t)$ (not shown) except in the vicinity of the origin, as above.  But the important difference with the previous case appears by looking at the $k$-integral function $\mathcal{H}(t)$, Fig.\ref{fig:kinteg}(b), which displays a surface below  the $k$-axis of the same order of magnitude as the one above the axis. This figure  suggests a three-step dynamics:  first  a global flip-flop motion (see the first positive peak  followed by a negative peak), then a period of  mixed transfer, then another flip-flop. This complex behavior is specified when looking at the $k$-profiles of  the test function $H$. The first flip-flop is illustrated in Fig.\ref{fig:testH-limit}(b) which shows  a direct transfer of energy from $k_{1}$ towards the whole band of modes (recall that this  band includes the frequencies up to the third harmonic of $k_{1}$)  followed by an inverse transfer of energy of this group of modes towards $k_{1}$. This riding and ebb tidal-like  motion is followed by a period of mixed transfer where  the  modes separate, the $k$-profiles having  positive and negative extrema, as illustrated by the red solid curve in Fig.\ref{fig:testH-limit}(c).  Then the negative amplitude part of the curve disappears, leading to a positive amplitude  profile, dashed curve in Fig.\ref{fig:testH-limit}(c), indicating that the transfer is mostly direct, this is followed by a decrease of the whole profile which becomes mostly negative (profile not shown), as  announced by Fig.\ref{fig:kinteg}(b).

In summary the  transfer of energy  from  a mode $k_{1}$ to the neighbors changes noticeably as $k_{1}$ approaches the  inertial/dissipative frontier. Here we did not observe a  clear cascade  towards small scales (with  transfer times increasing with $k$), but a complex flip-flop transfer concerning a large band of modes  which are first stuck together, and then separated. The  time duration of the first flip-flop exchange is $16$ms, it is also the time interval for the mode separation phase and for the subsequent direct transfer. We shall take it as the characteristic time for an exchange of energy near the $k_{1}$-mode. 

While  the two cases presented in this section display very different dynamics, one with a cascade process, the other one with a tide motion of the whole band of modes, it is interesting to compare their characteristic times, and to see if their ratio follows the scaling law given by equation (\ref{eq:taul}). In the inertial range, taking the time delay of the first direct transfer as characteristic of the transfer, we have $\tau= 32ms$ for  $k_{1}= 6182$. Here we have  $\tau '= 16ms$  for  $k'_{1} = 17300$ (same units).  The ratios are,
\begin{equation}
\frac{\tau}{\tau'}=2 \qquad   ( \frac{k_{1}}{k'_{1}})^{-2/3}=1.98
\mathrm{.}
\label{eq:sc}
\end{equation}
which is surprisingly  in prefect agreement with the scaling laws predicted by $K41$. 

In summary, our analysis of the test function $H$ allows to claim that the direct cascade does exist in the inertial range, but does not persist as one approaches the frontier with the dissipative range. When the spatial frequency of the donor mode is close to the boundary, the cascading direct transfer is replaced by a tide motion of a large band of modes including the second and third harmonics, followed by a complex behavior where direct cascade is absent.  This study could be used to  precise  where the frontier is lying.

\subsubsection{ Dissipative range in Modane data}
\begin{figure}
\centerline{
(a) \includegraphics[height=2.5in]{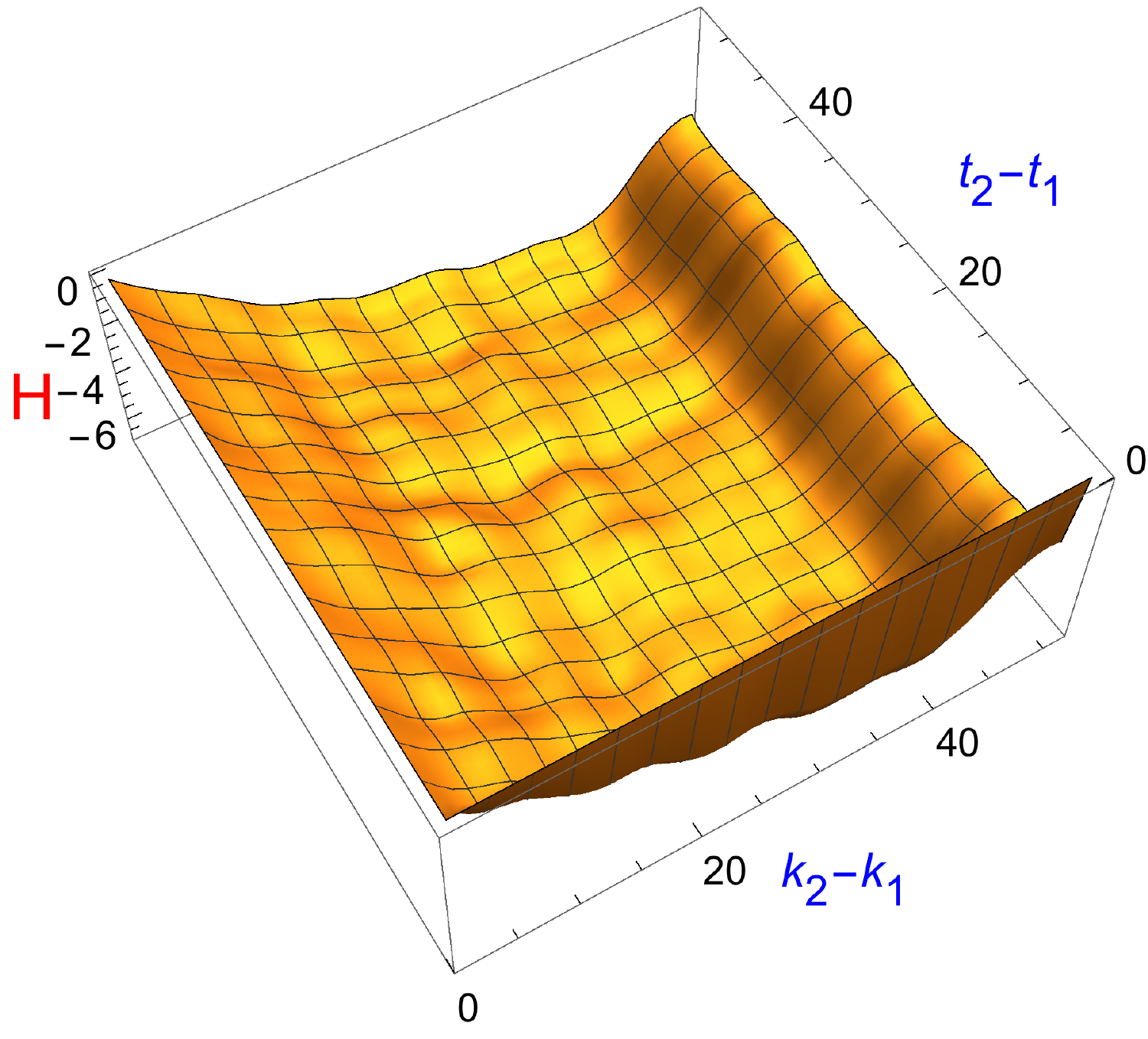}
(b)  \includegraphics[height=1.5in]{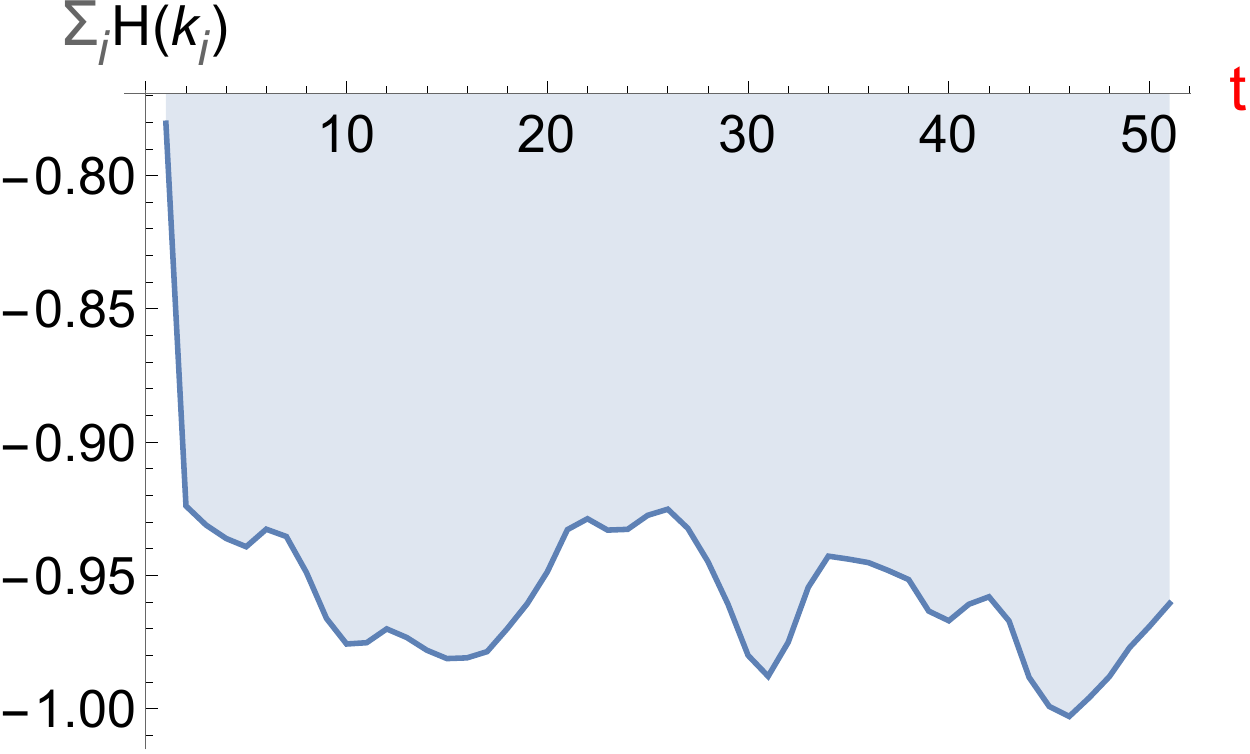}
 }
\caption{
$H$-test  for 
 Modane Eulerian velocity data with $k_{1,2}$ both in the dissipative range; $k_{1}=10^{5}$ in units of Fig.\ref{fig:spectre},  $k_{2, max}=2 k_{1}$ (a) three dimensional plot of the $H$ function; (b) $k$-integral function. Time is scaled to  $5 t_{s}$.
}
\label{fig:dissipH}
\end{figure}

Finally we have investigated the far-edge of the dissipative domain.  Fig.\ref{fig:dissipH} shows the test function for $k_{1}=10^{5}$  in units of Fig.\ref{fig:spectre}, $k_{2}$ extending up to the second harmonic, $2 \; k_{1}$  (the  limit  frequency of the spectrum is $3k_{1}$). The time range of this study is equal to $10$ms. Taking as reference the transfert time observed in Fig.\ref{fig:prof-50}, this time interval of $10$ms is twice larger than the characteristic time associated to the scalings of $K41$.

The figure shows that $H$ is negative in the whole range of $k$, as confirmed by the  $k$ profiles study which have  negative amplitudes. Therefore we put in evidence an  inverse transfer of energy towards $k_{1}$ (from all modes up to its second harmonic)  during a time interval twice longer than the characteristic time predicted by the $K41$ scaling law. One may infer that this first stage  is followed by more complex  dynamical behavior, as  suggested by the analysis of the preceding subsection, which also displays no direct cascade.

\section{ Wave turbulence}
\label{sec:wave-turb}
Wave turbulence can be observed in wavy systems where waves interact between each other through 
nonlinear interactions. When the amplitude of the waves are small, these nonlinear interactions can be
assumed small compared to the linear waves so that an asymptotic perturbation theory can be deduced, the so-called weak wave turbulence theory (WWT further on). First developed about fifty years ago for water and plasma waves~\cite{hasselmann,benney,zakgrav66,zakplasma67,zakcap67}, it leads to a 
kinetic equation for a quantity linked to the energy spectrum.
Beside the equipartition of the energy, another stationary solution of this equation
exists corresponding in general to a constant flux of energy from large to
small scales, called the Kolmogorov-Zakharov (KZ) spectra
~\cite{ZakhBook,NewellRumpf,NazBook}. Since then, wave turbulence has been investigated and observed in many physical systems, from the initial water waves problems~\cite{Falcon07,Falcon09} to nonlinear optics~\cite{Dyachenko-92}, Alfven waves~\cite{Galtier} and recently elastic waves in plates~\cite{during,arezki,mordant08}. In the present work, we will use a simplified 1D model of wave turbulence to investigate irreversibility process, allowing rapid and extensive numerical simulations.

\subsection{A prototype model of wave turbulence}
We use one of the so-called MMT (for Majda-MacLaughlin-Tabak) model equations first introduced in~\cite{MMT} that are deduced from the Non Linear Schr\"{o}dinger equation (NLS) in order to obtain dispersive waves features for which wave turbulence holds. It reads for the complex function $\psi(x,t)$:

\begin{equation}
\imath \partial_t  \psi=|\partial_x|^{1/2} \psi+ |\psi|^2 \psi
\label{mmt}
\end{equation}

where the linear operator $|\partial_x|^{1/2}$ corresponds to the  fractional derivative $(-\partial_{xx})^{1/4}$ (named Riemann-Liouville derivative), which is easily
defined in the Fourier space by the coefficient $\sqrt{|k|}$. The function $\psi$ is often called the
wave function by analogy with the NLS context and can be written in term of its density $\rho$ and phase $\phi$ through the transform $\psi =\sqrt{\rho} e^{i\phi}$. Linear waves ($\propto e^{i kx-\omega t}$) 
obey thus the following dispersion relation $ \omega =\sqrt{|k|}$ that is similar to the one of the gravity 
waves in deep ocean, so that this model can be slightly considered as a 1D analogy of water wave 
dynamics. Moreover, it removes the integrability property of the usual NLS equation in one space 
dimension. The choices of this dispersion relation and of the specific nonlinear term inside the general
framework of the MMT models have also been dictated by the need of exhibiting wave turbulence in 
the dynamics. MMT equation have been investigated originally to test the validity of
the WWT theory~\cite{MMT}, showing differences in some cases that have been explained by the detailed study of the dynamics~\cite{dias}. We are not concerned in our work by these specific issues since we have chosen a version of the MMT equations that exhibits WWT solutions at low forcing.

The dynamics of the MMT equation conserved two integral quantities, 
the mass or number of particles $N$, defined by:

\begin{equation}
N= \int |\psi|^2 dx.
\label{masse}
\end{equation}

and it obeys also a Hamiltonian dynamics, yielding:

$$ \imath \partial_t \psi= \frac{\delta {\cal H}_a}{\delta \psi^*} $$

where $\psi^*$ is the complex conjugate of $\psi$.
${\cal H}_a$ is the Hamiltonian, conserved by the dynamics and defined by
\begin{equation}
{\cal H}_a=\int \left( \left ||\partial_x|^{1/4} \psi \right|^2+\frac{1}{2} |\psi|^4 \right)dx.
\label{hamilton}
\end{equation}

The first term of the Hamiltonian is called the kinetic energy (noted ${\cal E}_c$ later on), while the other one corresponds to a nonlinear interaction potential.

The usual WWT theory has been applied to this MMT model leading to a kinetic equation for the
density spectrum $n_k$ defined by: 
$$ n_k=<|\psi_k|^2>, $$
where $\psi_k=\int \psi e^{ikx}dx$ is the Fourier transform of $\psi$. Notice that the kinetic energy density writes in Fourier space simply:
$$ {\cal E}_k = \sqrt{|k|} |\psi_k|^2.$$
The kinetic equation is deduced under the assumption of small nonlinearity so that an asymptotic perturbative 
expansion can be deduced, leading to the kinetic equation. More precisely, this approach assumes that the 
nonlinear term is smaller than the linear term so that it can be treated as a correction to the linear equation, 
indicating eventually that the amplitude of the modes is small, roughly speaking $<|\psi|^2> \ll 1$. The kinetic 
equation describes the nonlinear interaction between the wave through a four waves resonance process~\cite{Dyachenko-92}.
This kinetic equation exhibits four power law type stationary solutions: two equilibrium ones, corresponding to the equipartition of the two conserved quantities, namely the mass yielding $n_k \propto 1$ and the Rayleigh-Jeans spectrum for the equipartition of the kinetic energy $n_k \propto 1/\omega \sim k^{-1/2}$. 
As shown firstly by Zakharov for plasma waves~\cite{zakplasma67}, two other stationary solutions to the
kinetic solution exist, associated to the flows of the two conserved quantities through scales, namely:
\begin{equation}
n_k \propto P^{1/3} k^{-1}
\label{KZspec}
\end{equation}
describing the direct cascade of the constant energy flux $P$ from large scales (small $k$) to
small scales (large $k$).
Furthermore, the inverse cascade of mass $Q$, from small to large scales, is also present, yielding:

\begin{equation}
n_k \propto Q^{1/3} k^{-5/6}.
\label{KZspec}
\end{equation}

In the following, we will thus use this equation to characterize irreversibility numerically in the direct cascade of energy configuration, obtained by forcing the system at large scale and damping it at
small scale, with an additional pumping at large scale to avoid accumulation of mass at large scales.

\subsection{Direct cascade of energy}

To investigate numerically the direct cascade of energy in the MMT equation (\ref{mmt}), we add forcing and pumping at large scale and damping at small scale, solving the following equation:

\begin{equation}
\imath \partial_t \psi=|\partial_x|^{1/2} \psi+ |\psi|^2 \psi +\imath ({\cal I}-{\cal D}-{\cal P}) 
\label{mmtnum}
\end{equation}
where ${\cal I}$, ${\cal D}$ and ${\cal P}$ are the injection, dissipation and pumping term respectively.
To mimic large scale injection and pumping and small scale dissipation, we define them in the Fourier space, yielding:
\begin{equation}
{\cal I}_k=A \Theta(k,t) \,{\rm for}\, |k|<k_i; \,\, {\cal D}_k=\alpha (\sqrt{|k|}-\sqrt{k_c}))\psi_k \,{\rm for}\, |k|>k_c; \,\, {\cal P}=-\alpha \psi_k \delta(k)
\end{equation}
where $A$ is the amplitude of the forcing, $\alpha$ the damping coefficient that we take equal to the pumping one, which is acting only on the $k=0$ mode. $\Theta(k,t)$ is a delta correlated zero mean white noise that is 
computed using the random functions of the computer. It is important to emphasize here that the forcing, 
although imposed at large scale, does not only involve energy injection since both energy and mass are coupled.
Therefore, this forcing has to be interpreted as an injection of both mass and energy at large scale and the
WTT predicts that most of the energy should cascade towards small scales (taking with it a small amount
of mass) while most of the mass will transfer towards even larger scales where it will be pumped.

\begin{figure}
\centerline{
\includegraphics[width=4in]{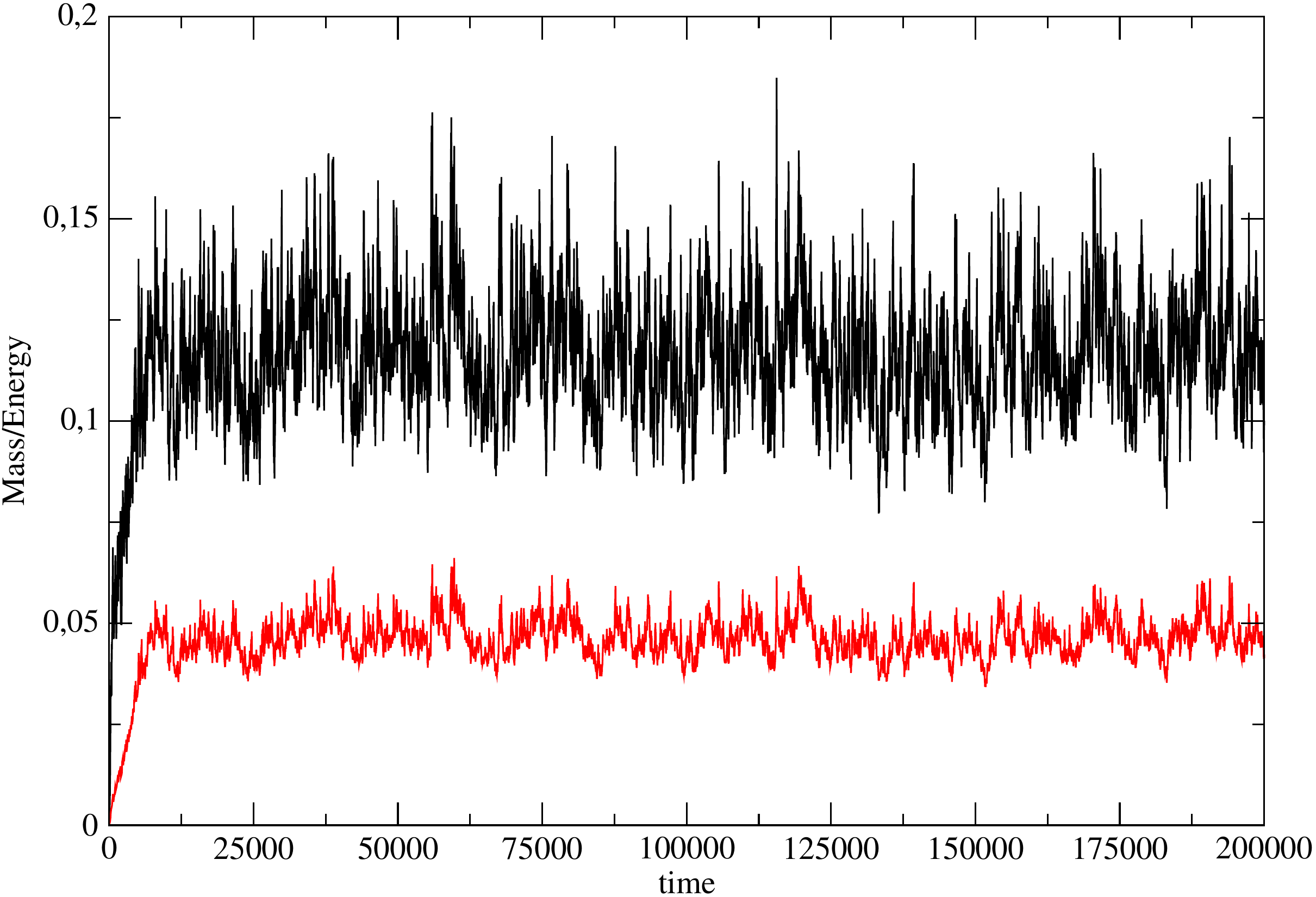}
  }
\caption{The mass $N(t)$ (top black curve) and the energy ${\cal H}_a(t)$ (red bottom curve) as function of time for the numerical simulation of the MMT equation (\ref{mmtnum}) with forcing, pumping and dissipation. After a short transient of about $10000$ unit time a statistically stationary regime is reached where both the mass and the energy fluctuate around a mean value. The simulation was performed on $8192$ grid points using $dx=0.5$ so that the length of the system is $L=4096$. The amplitude of the forcing is $A=0.01$ with the injection scale $k_i=0.005$ while the dissipative scale is $k_c=\pi$ and the dissipation coefficient is $\alpha=0.1$. The time step is $dt=0.05$. Only a zoom at small time is shown here, the simulation being performed until $t_{max}=5. \cdot 10^6$ time units.}
\label{massener}
\end{figure}

\begin{figure}
\centerline{
a)\includegraphics[width=3in]{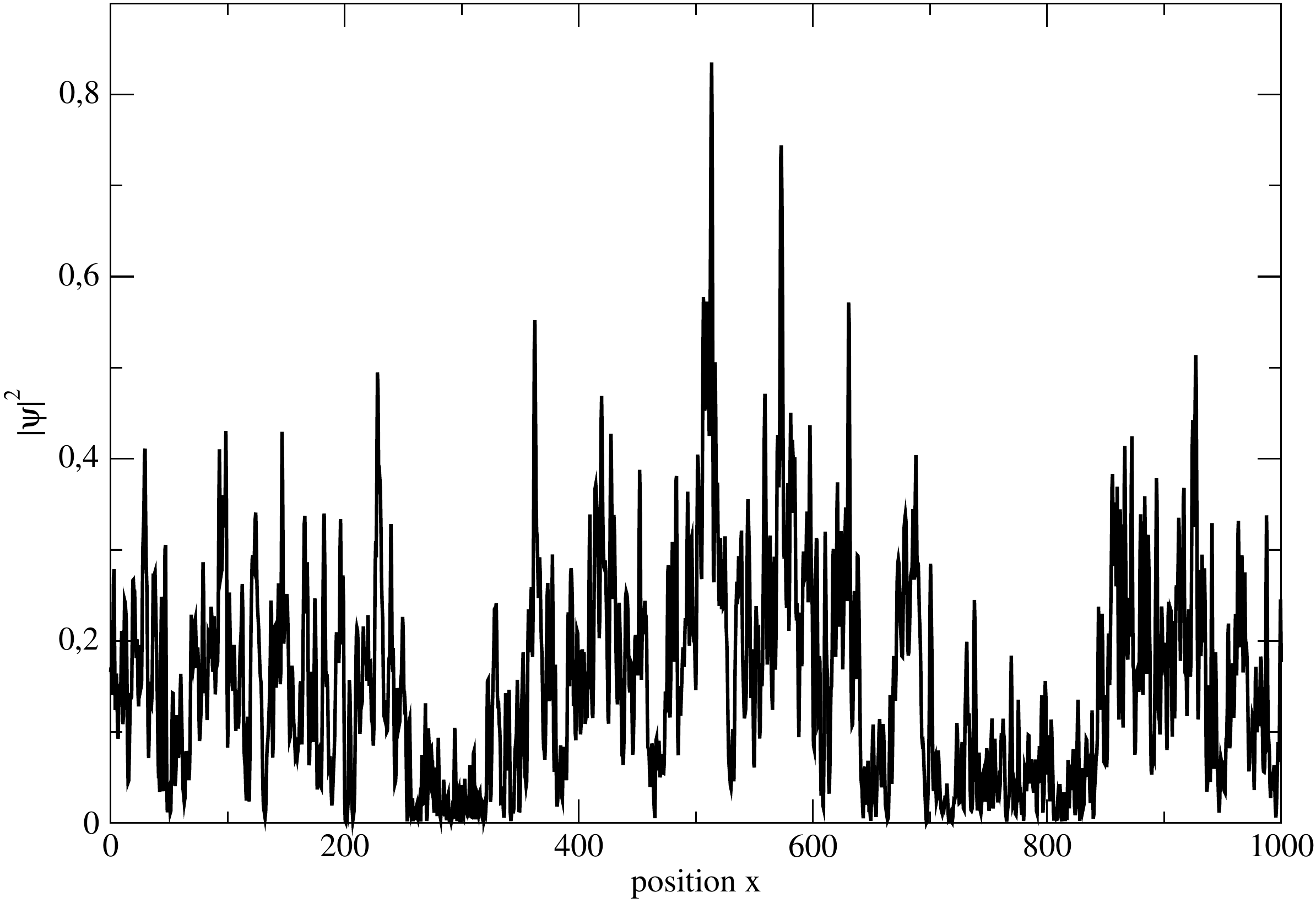} b)\includegraphics[width=3in]{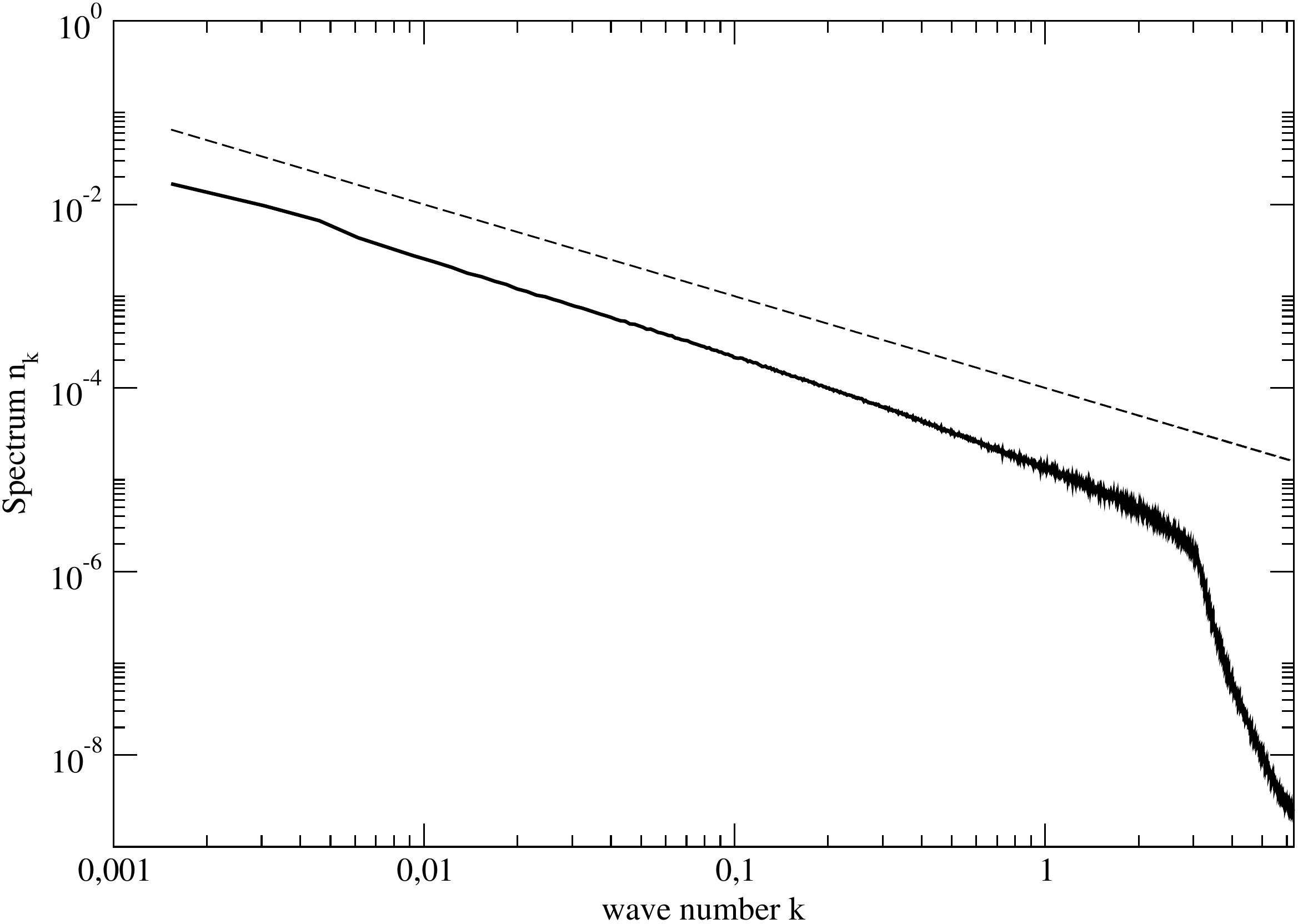}
}
\caption{a) Typical density field $|\psi(x)|^2$ of the solution of the MMT equation in the stationary regime for the same simulation than that of Fig. \ref{massener}. Only a quarter of the total domain is shown for the sake of visibility. b)Spectrum $n_k=<|\psi_k|^2>$ as a function of the wave number $k$. The statistical average is computed in the stationary regime as the mean over $4000$ instantaneous spectra. The slope $k^{-1}$ corresponding to the expected WWT turbulent spectra of direct energy cascade is shown as a dashed line for eye-guiding.}
\label{spectrum}
\end{figure}

The MMT equation is numerically simulated using a pseudo-spectral method where the kinetic term in solved in the Fourier space where it is simply a rotation, while the nonlinear term is solved in real space where it is again a 
rotation. This process conserves exactly the mass, while the injection, pumping and dissipation are
added in the Fourier space (these terms makes that the final equation obviously does not conserve neither the mass nor the energy). Starting with a null field everywhere $\psi(x,0)=0$, a steady regime is reached after some transient, where both the total mass $N(t)$ and the energy ${\cal H}_a(t)$ fluctuate
around some constant values, as shown on Fig. \ref{massener}. In this regime, the injection is balanced by the dissipation at short scales and the pumping at large scales, reaching an out of equilibrium steady state typical of turbulent dynamics. In this regime, the spectrum follows the WWT prediction, as shown on Fig. \ref{spectrum}, where the spectrum is computed using a time averaging over a large time window. For all the figures shown in this paper, the same values of the numerical parameters have been used, namely $dx=0.5$ on
 $N=8192$ grid points, so that the total length of the system is $L=4096$. The injection amplitude is $A=0.01$ with $k_i=0.05$. The dissipation and pumping coefficient is $\alpha=0.1$ using $k_c=\pi$. The time step for
 the simulation os $dt=0.05$ time unit.

\subsection{Irreversibility global test functions}

\begin{figure}
\centerline{
\includegraphics[width=4in]{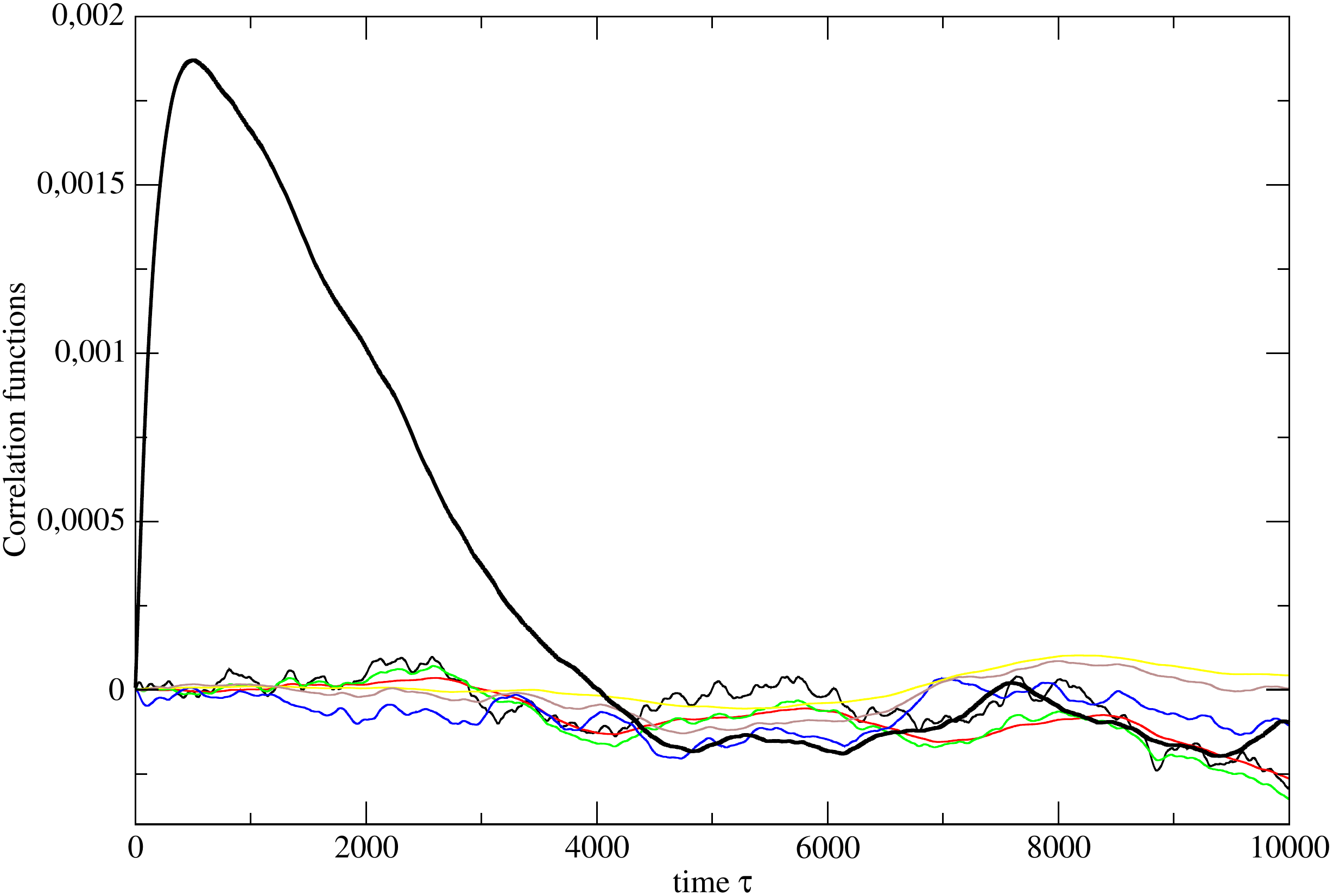}
  }
\caption{(Color online) The correlations functions $\Psi_1(\tau)$ and $\Psi_2(\tau)$ using both $x=(N(t)-<N>)/<N>$ and $x=({\cal H}(t)-<{\cal H}>)/<{\cal H}>$ and the cross-correlation function $\chi(\tau)$ as function of the time delay $\tau$. All the functions $\Psi_i$ are below $10^{-4}$ (indistinguishable in the figure), indicating no 
irreversible signal both for the mass and energy fluctuations. On the other hand the cross-correlation function 
$\chi$ indicates a positive correlation between the mass and the energy.}
\label{fluctglob}
\end{figure}

It is first tempting to investigate if irreversibility can be characterized directly from the statistical fluctuations of the total mass and the kinetic energy of the system, using the test functions $\Psi_1$ and $\Psi_2$ introduced above both for the normalized quantities $x=(N(t)-<N>)/<N>$ and $x=({\cal E}_c(t)-<{\cal E}_c>)/<{\cal E}_c>$, 
leading to the function $\Psi_i^{n}$ and $\Psi_i^{h}$ respectively, for $i=1,2$. As shown on Fig. \ref{fluctglob}, these function remain very small (typically below $10^{-4}$) indicating that there is no or only little irreversibility in the mean mass and mean energy fluctuations. However this test yields much clearer results when one
considers the cross correlation function of these two quantities, namely:
\begin{equation}
\chi(\tau)=\frac{<N(t){\cal E}_c(t+\tau)>-<{\cal E}_c(t)N(t+\tau)>}{<N><{\cal E}_c>}
\label{correl}
\end{equation}
which is also plotted on the same figure and that exhibits a clear positive behavior for small enough correlation time $\tau$, two orders of magnitude higher than the fluctuations behaviors observed for the $\Psi_i$ functions. It clearly indicates that the mass and the energy of the systems are {\it irreversibly} correlated showing the irreversible transfer of the mass injected with the energy later on. The interpretation of such a correlation is not obvious, although 
one could argue that it comes from the injection at large scale that involves both mass and energy. More 
precisely, the mass and energy injected at large scale exhibit a different dynamics due to the direct cascade of
energy: while the mass is rapidly transferred and then pumped at $k=0$, while the energy remains longer in the system through the direct 
cascade process, leading to the positive cross-correlation.
Finally, it is important to notice that this function behaves at short times linearly in $\tau$, so that singularities in the time derivative of the fields is expected in the dynamics, as 
explained for the Modane experiments.

\subsection{Test on the spectrum correlation}

In the same vein than what has been shown for the Modane experiment above, we will investigate
similar time dependent cross-correlation functions $H(k_1,k_2,t_2-t_1)$, 
using the former result that mass and kinetic energy dynamics are irreversibly coupled.
More precisely, since the mass spectrum is 
$$ N_k(t)=|\psi_k|^2$$
while the kinetic energy spectra reads:

$$ {\cal E}_k(t)=\sqrt{|k|} |\psi_k|^2,$$

we will consider the cross-correlation functions on the fluctuations of these quantities:

 \begin{equation}
H (k_{1},k_{2}, t)= <(N_{k_{1}}(t_{1})-<N_{k_1}>) ({\cal E}_{k_{2}}( t_{2})-<{\cal E}_{k_{2}}>)>  - <(N_{k_{2}}(t_{2})-<N_{k_2}>) ({\cal E}_{k_{2}}( t_{1})-<{\cal E}_{k_{1}}>)>
\mathrm{.}
\label{eq:testHWTT}
\end{equation}

Again, we will observe that such functions exhibit irreversible features of the dynamics indicated by the positive value of  the function, although the signal is often noisy and oscillating with time $\tau$ and wave numbers. 
We will here not reproduce the same detailed analysis than 
for the Modane experiments but only show the 3D displays of the function $H$ in the three regimes identified 
above: the inertial, the transition to dissipation and the dissipative regimes. Only a qualitative discussions of the results is provided here and we postpone a more detailed analysis of these datas to further works. In particular,
since the WWT involves four waves resonance, it would be interesting to investigate the correlation functions
for resonant subset of wave numbers~\cite{Aubourg15}.

\subsubsection{In the inertial range}

\begin{figure}
\centerline{
\includegraphics[angle=270,width=3.5in]{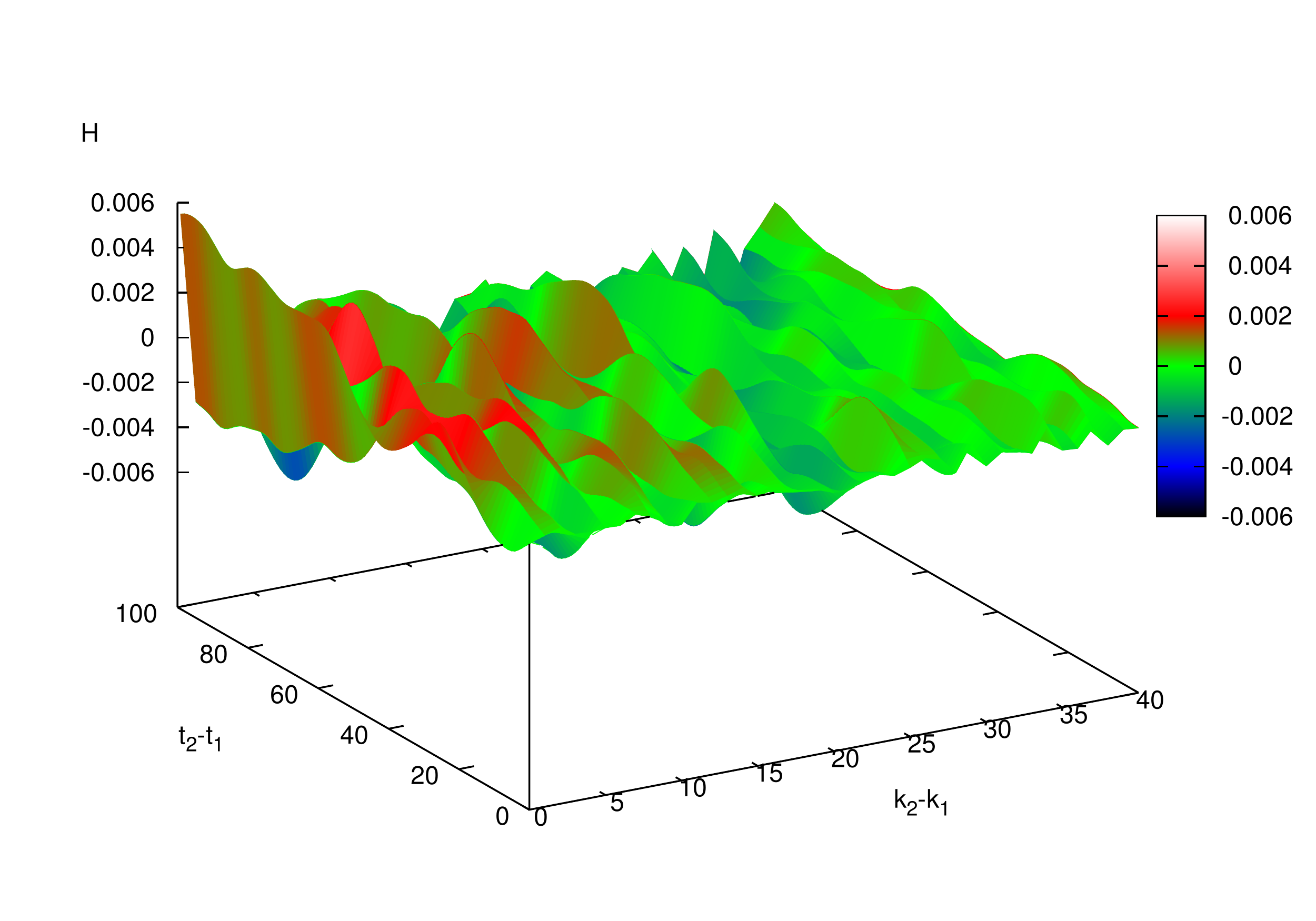}
}
\caption{
Test  function $H (k_{1},k_{2}, t_{2}-t_{1})$  computed from the MMT equation in the stationary regime, for $k_{1},k_{2}$ in the inertial range. In units of Fig.\ref{fig:spectre} $k_{1}=200\pi/L$, and $k_{2}$ extends up to $3 k_{1}$. The $\tau=t_2-t_1$ time axis is in time units.
}
\label{iner3D}
\end{figure}

We thus show on Fig. \ref{iner3D} the 3D plots of the function $H$ for wave numbers $k_1$ and $k_2$ in the inertial range. A clear positive correlation of the signal is observed, that develops at larger $k$ for larger time, 
indicating roughly the direct cascade of energy process.

\subsubsection{Transition domain between the inertial and the dissipative range}

\begin{figure}
\centerline{
\includegraphics[angle=270,width=3.5in]{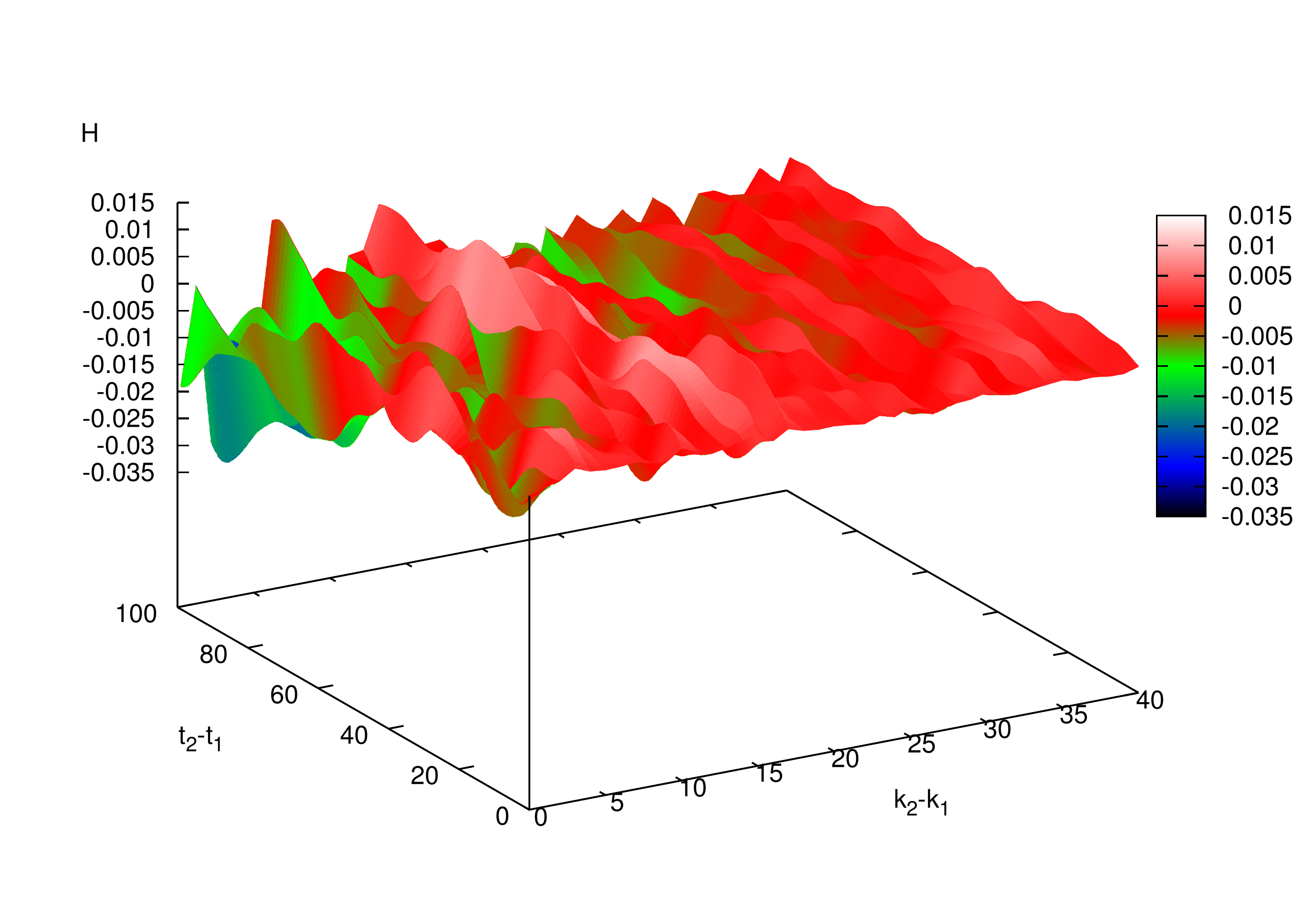}
}
\caption{Test  function $H (k_{1},k_{2}, t_{2}-t_{1})$  computed from the MMT equation in the stationary regime, for $k_{1}$ in the inertial range just before the dissipative range while $k_2$ is mostly in the dissipative range. In units of Fig.\ref{fig:spectre} $k_{1}=4000\pi/L=125\pi/128  \lesssim k_c=\pi$, and $k_{2}$ extends up to $4400\pi/L$. The $\tau=t_2-t_1$ time axis is in time units.
}
\label{trans3D}
\end{figure}

When $k_1$ is taken just before the dissipative range, the correlation function exhibits a different behavior, as shown in Fig. \ref{trans3D}.
While positive correlations can still be seen for short $\Delta k=k_2-k_1$, they almost vanish as $k_2$ 
penetrates inside the dissipative range.

\subsubsection{Dissipative range}

\begin{figure}
\centerline{
\includegraphics[angle=270,width=3.5in]{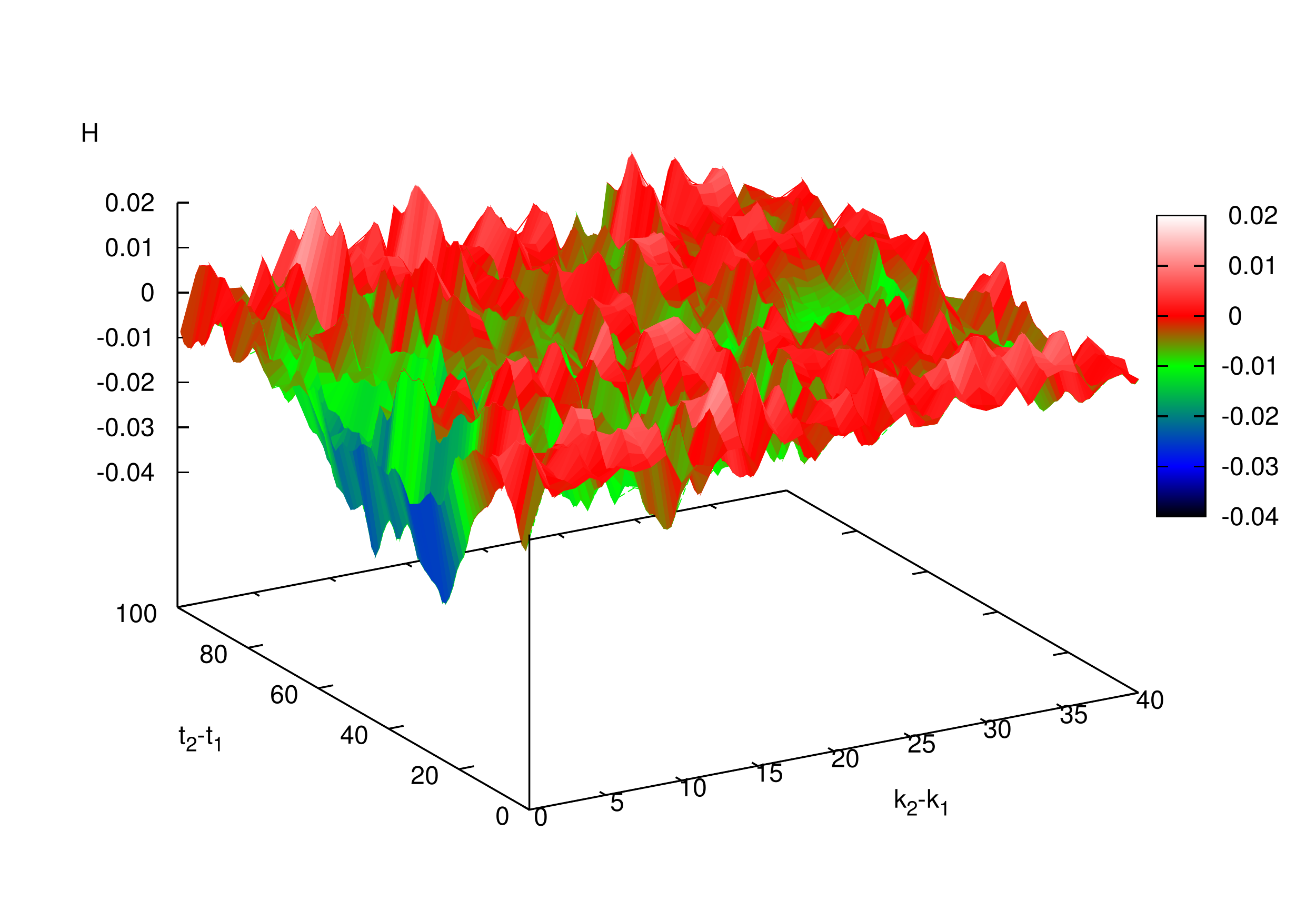}
}
\caption{Test  function $H (k_{1},k_{2}, t_{2}-t_{1})$  computed from the MMT equation in the stationary regime, for $k_{1}$ and $k_2$ deep inside the dissipative range. In units of Fig.\ref{fig:spectre} $k_{1}=6000\pi/L=375\pi/256 \> k_c=\pi$, and $k_{2}$ extends up to $6400\pi/L$. The $\tau=t_2-t_1$ time axis is in time units.
}
\label{diss3D}
\end{figure}

Finally, in the dissipative range, the correlation signal appear almost as a fluctuation field (see Fig. \ref{diss3D}), with large negative domains. There, no more
coupling is expected from the mass and energy fluctuations,.

\section{Conclusion and perspectives}
\label{sec:concl}
A fundamental difficulty in theory of fully developed turbulence is the non existence of a Gibbs-Boltzmann like probability for the fluctuations. At equilibrium there is no need to solve the complex dynamical equations of the many body problem to derive the statistics of the equilibrium fluctuations. On the contrary there is no known way to get the probability distribution of fluctuations of the velocity and other quantities in a turbulent flow. Likely one fundamental difference between equilibrium and non equilibrium systems is the breaking of time reversal symmetry in such non equilibrium systems. We did show in this paper that this breaking shows up by analyzing the time correlations of the Eulerian data measured in the Modane wind tunnel with its very large Reynolds number and the data derived from the integration of a partial
differential equation (a model of wave turbulence), which also shows transfer of energy across the wave number
spectrum. This analysis yields a number of results, some expected others not. As had been predicted in ref \cite{pom82}, there is a clear signal in the correlation functions showing that the energy is transferred from large scale to smaller scale with a time delay. This result shows that the cascade is not only an image for describing the spatial dependence of velocity fluctuations in a turbulent flow but also a way to describe the dynamics occurring in such a turbulent flow. But, by analyzing the test function for irreversibility we found that, contrary to expected, this test function is linear with respect to time for small time differences whereas a straightforward calculation yields a dependence proportional to the time cube. Because the coefficient of the term linear with respect to time is (formally) proportional to the Eulerian acceleration, we attempted to relate the anomaly in the small time expansion of $ \Psi_2(\tau)$ near  $\tau = 0$ to an anomaly in the statistics of the large deviations of this acceleration. A direct analysis of the statistics of the acceleration confirms this slow decay of the probability for large fluctuations. In the turbulence literature this kind of behavior is usually referred to as resulting from  a form or another of intermittency. This kind of explanation refers to statistical properties which are very hard to check directly. We believe that another kind of explanation is by the occurrence of finite time  singularities in solutions of the Euler equations. Supposing that the singular solutions are of the self-similar type one finds how the random passage of singularities near the detector changes two properties, the behavior of  $ \Psi_2(\tau)$ near  $\tau = 0$ and the statistics of large accelerations. This last property is known with  remarkable accuracy from the data and it turns out that it cannot be represented by a self-similar singularities with a single exponent.  

This led us to suggest that the impact of finite time singularities on the expression of various functions $ f(r,t)$ of space and time close to the singularity  is more complex  that the one usually assumed, based on the hypothesis of a simple self-similar solution with a unique exponent. We suggest to introduce an additional time dependent variable, in the spirit of Appell's work \cite{appell}  already generalized in \cite{CRASYP, CRAS-PP}.

The general idea on fully developed turbulence derived from our analysis is that first there is a cascade very much as described by Kolmogorov with a clear time dependence of the energy transfer with respect to the scale. However this does not seem to exclude the existence of other phenomena like the occurrence of finite time singularities which do not play a role in the transfer of energy from scale to scale, but which show up in the time records at a point of measure. This idea goes against a tight connection between intermittency as observed since a long time in turbulent flows and fluctuations in Kolmogorov cascade of energy. If, as we observe, the exponent derived from the behavior of   $\Psi_2(\tau)$ near  $\tau = 0$ is incompatible with a Sedov-Taylor scaling of the singularity, this singularity does not dissipate (once it is stopped by viscosity) any energy. Therefore it cannot be the end point of a cascade of energy with fluctuations in the energy transfer as resulting from the $K62$ theory.  

\bigskip
{\bf{Acknowledgments} }
We  thank Dr Yves Gagne for having provided us the  Eulerian  velocity  data of Modane?s experiment, which were taken  by him  et his collaborators, and we thank ONERA  for facility.

{}

 \end{document}